\documentclass[12pt,twoside]{article}
\usepackage{chadstyle}

\usepackage{booktabs}
\usepackage{tabularx}
  \newcolumntype{C}{>{\centering\arraybackslash}X}
\usepackage{multirow}
\usepackage{changepage}
\usepackage{enumitem}
\usepackage{placeins}    
\usepackage{float}       

\hypersetup{allcolors=black}

\begin{document}
\bibliographystyle{aernobold}


\begin{spacing}{0.9}
\begin{titlepage}

\title{Micro and Macro Perspectives \\
on Production-Based Markups}
\runningheads{Fernald, Gandhi, Ruzic, Traina}{Production-Based Markups}

\author{\large
    \href{https://www.insead.edu/faculty-research/faculty/john-fernald}{John Fernald},
    \href{https://amitgandhi.com}{Amit Gandhi},
    \href{https://www.insead.edu/faculty-research/faculty/dimitrije-ruzic}{Dimitrije Ruzic},
    and \href{https://james-traina.com}{James Traina}\thanks{Fernald: john.fernald@insead.edu. Gandhi: amitgandhi@gmail.com. Ruzic: dimitrije.ruzic@insead.edu. Traina: james.traina@nyu.edu. We thank Susanto Basu, Arshia Hashemi, seminar participants at INSEAD, the Federal Reserve Banks of San Francisco and Richmond, and the University of Arizona, as well as our editor David Romer and several excellent anonymous referees for helpful comments and feedback.}
   }

\date{\small April 2026}
\maketitle
\thispagestyle{empty}

\vspace{-0.3in}

\begin{abstract}
We review the ``production approach'' to estimating markups, the ratio of price to marginal cost. The approach is uniquely scalable: it requires no model of consumer demand or market structure and applies broadly across firms, industries, and time. Our organizing insight is that the production-based markup is a residual. Like the Solow residual, it is clean in theory but potentially contaminated by misspecification and mismeasurement. This framing helps explain why small differences in implementation can produce starkly different results from the same data. In some cases, markups have risen sharply. In others, they have not. Despite the disagreements in the literature, the importance of understanding and measuring market power cannot be overstated. We provide conceptual rationales for this disagreement, offer practical guidance on data and estimation, and call for greater transparency about how much of the variation attributed to markups may instead reflect technology.
\end{abstract}

{\medskip \footnotesize
    JEL Codes: D24, D43, E22, E23, L11, L16, O33, O47
    }

\end{titlepage}
\end{spacing}

\section{Introduction}

We review the ``production approach'' to estimating markups. Intuitively, markups are about pricing; pricing relates to marginal cost, and marginal cost reflects the production constraints facing firms. The increased availability of firm-level microdata and new econometric methods makes it easier to estimate production functions and infer markups. Our review focuses on this recent firm-level literature, with a goal of bridging micro and macro perspectives. Questions of market power matter for many fields, from industrial organization (IO) to international trade to macroeconomics. We highlight the challenges in moving from firm-level analyses to industry and economy-wide conclusions.

\citet{HallBPEA1986, Hall1988relation, HallR90} introduced the production approach. He showed that cost minimization implies the markup $\mu$ can be written in a remarkably simple way as the ratio of a flexible input's output elasticity $\gamma$ to the cost of that input as a share of total revenue, $s$. That is, $\mu = \gamma/s$. Markups can be read directly from producer behavior. It is not necessary to specify the demand system or market structure. This makes it attractive compared to typical methods from IO that are tailored to narrow industries. The production approach scales to large panels of firms across the economy and over time.

Hall and a large body of work that followed apply his insight to industry data in growth rates. \citet{deloecker2012markups} highlight challenges in applying Hall's industry approach to firm-level data. They propose a variant of Hall's method that focuses on a single input and uses modern methods for estimating firm-level production functions. A selling point is that the new approach allows estimation of firm-by-firm markups.

In recent years, interest in market power has soared. The interest reflects in part that markups may have risen sharply, and in part that rising markups might provide a unified explanation for a range of macroeconomic puzzles \citep{deloecker2020rise}. These include declining labor shares, rising economic profits, slow productivity growth, weak investment, and declining dynamism.\footnote{References to these puzzles include, respectively, \citet{elsby2013decline}, \citet{karabarbounis2014global}; \citet{barkai2020declining}, \citet{karabarbounis2019accounting}; \citet{FernaldInklaarRuzic2024}, \citet{gutierrez2017investmentless}, \citet{crouzet2019understanding}; and \citet{davis2007volatility}. \citet{Syverson2025_ARE} provides a complementary survey covering both markups and markdowns.} 

More broadly, modern macroeconomics rests on a foundation of imperfect competition. In endogenous growth models, firms charge a markup to cover innovation costs \citep{romer1990endogenous}. In New Keynesian models, firms are monopolistically competitive price-setters, not perfectly competitive price-takers \citep{RotembergWoodford1999}. Firm-level markup estimates, applied at scale, allow economists to incorporate rich heterogeneity, strengthening the microfoundations of macro analysis. IO and trade increasingly need the same scalability: IO is moving from individual markets to cross-industry patterns \citep{dopper2025rising}, and trade needs markup estimates across countries and industries \citep{FeenstraWeinstein2017}.

However, the empirical evidence for rising markups is far from settled. Some production-approach specifications (such as \citealt{deloecker2020rise}) show dramatic increases in recent decades. But others do not (such as \citealt{FosterHaltiwangerTuttle2024, traina2018aggregate, BenkardMillerYurukoglu2025}).


To put the empirical disagreements in context, our review emphasizes that production-based markups are a residual. Using the equilibrium first-order condition for a fully flexible input, the markup absorbs the entire gap between the output elasticity and the revenue share. Unlike a regression error or a structural demand shock, this residual is itself the object of interest: an equilibrium ratio that holds under any model of competition. Three things can go wrong: wrong first-order condition, wrong output elasticity, or wrong revenue share. All three deserve scrutiny.

The residual framing is analogous to the \cite{solow1957technical} residual. The Solow residual absorbs all output growth not explained by input growth. Under standard conditions, it is a clean measure of technology change. But just as a large literature shows that non-technological factors contaminate the Solow residual as a measure of technology,\footnote{Non-technological factors in the Solow residual include markups \citep{HallBPEA1986, Hall1988relation, HallR90}, factor utilization \citep{BFK2006technology, fernald_quarterly}, aggregation \citep{basu1997returns, baqaee2020productivity}, and mismeasurement \citep{corrado2009intangible, byrne2016does}.} (mis)measurement and (mis)specification can contaminate production-based markups as a measure of market power. Differences across papers---even those using the same data---can therefore yield different residuals. Contamination is the price of scalability: the approach avoids specifying demand or market structure, but what it omits shows up in the markup.

Even if we resolve all the empirical disagreements, the implications of market power depend on its cause. The production approach we introduce in Section \ref{s:approach} is a measurement tool; it is silent on structural drivers. The first-order condition is an equilibrium relationship. It holds regardless of what generates the markup, so demand and market structure need not be specified. But for some questions, we need the structure. For example, are rising markups ``bad,'' reflecting barriers to entry, or ``good,'' reflecting returns to innovation? Answering that question requires structure that the production approach was never designed to provide. The IO literature offers that structure for narrow markets. The production approach offers breadth. Both perspectives are needed. We return to these issues at the end of the paper.

Unfortunately, the devil is in the implementation. Implementing the production approach requires substantive choices, each of which layers on different auxiliary assumptions. The ``garden of forking paths'' \citep{Borges1941_garden, GelmanLoken2014} in Section \ref{s:garden} documents the sensitivity across two (of multiple) such choices. One is which input to treat as flexible, usually labor or materials. Another is how much variation to allow in output elasticities across firms and over time. Several studies use the same data and estimation techniques to show these choices matter. The results are stark. Some specifications imply sharply rising markups; others do not.

The remainder of the review examines what we know about the sources of this sensitivity. 

Section~\ref{s:conceptual} asks whether the source of empirical disagreements is conceptual. The first-order condition may be missing wedges that contaminate the markup estimate, including adjustment costs, monopsony power, or bargaining frictions that drive a gap between observed input prices and the allocative costs to which firms actually respond. The production function may also allow too little variation in output elasticities, forcing genuine technological heterogeneity into the markup residual instead. A recurring theme is the tension between markups and technology: assumptions that restrict elasticity variation load that variation onto the markup. A related conceptual fork, also taken up in Section~\ref{s:conceptual}, is the choice between gross-output and value-added production functions. With markups, both the theoretical and the empirical case for value-added production functions is weak \citep{BasuFernald1995, gandhi2017heterogeneous, Ruzic:2024}.

Sections \ref{s:data} and \ref{s:econometrics} examine the interplay of data and estimation that researchers implementing the production approach need to understand. Section \ref{s:data} focuses on the data constraint: missing inputs, misclassification of variable and fixed costs, and limited sample coverage can all distort revenue shares or bias estimated elasticities. Section \ref{s:econometrics} examines the econometric challenge of recovering output elasticities when inputs respond to unobserved productivity and prices. Estimation in this setting is harder than often recognized because the standard toolkit was largely built for perfect competition and better data than what we typically have.

Despite these challenges, the importance of understanding and  measuring market power cannot be overstated. Markups shape resource allocation, welfare, and how shocks propagate across firms and industries---and no other method estimates them at comparable scale. The difficulty of the measurement problem is not a reason to set the production approach aside, but a reason to invest in improving it.

Therefore, Section~\ref{s:Call_to_Arms} closes with three calls to arms, all centered on the tension between market power and technology. First, we call for transparency: systematic reporting of how much revenue-share variation is explained by markups versus output elasticities. Second, we call for testing: comparing production-based markups against demand-based estimates from IO, quasi-experimental evidence, and simulations. Third, we call for more work mapping firm-level markup and technological heterogeneity into macroeconomic models. Getting the implementation right matters precisely because the approach is so broadly applicable.

We now turn to laying out the production approach at the heart of this review.

\section{The production approach to markup estimation} \label{s:approach}
Production functions and markups are inextricably linked through factor demand. Intuitively, firms with market power raise prices and reduce output, which in turn leads them to use fewer inputs. \citet{HallBPEA1986, Hall1988relation, HallR90} showed how to use this insight to estimate markups assuming little other than cost minimization. The rest of the firm's profit maximization problem is not needed.

The first four subsections present the main conceptual material. Subsection~\ref{s:unifying_FOC} presents the key cost-minimizing first-order condition (FOC). Subsection~\ref{s:growth_rates} describes how Hall and others implemented this approach using data in growth rates, emphasizing three lessons that remain relevant for the newer microdata literature. Subsection~\ref{s:new_literature} introduces that newer literature, which applies the FOC to a single input and usually estimates production relationships in levels. Subsection~\ref{s:implementation} previews the key conceptual choices that drive the empirical disagreements documented in the rest of the review.

\subsection{The unifying first-order condition}\label{s:unifying_FOC}
Consider a production function for firm $i$ in period $t$:
\begin{equation}\label{e:production_function_general}
Y_{it} = F(X^1_{it}, \ldots, X^N_{it}, A_{it}),
\end{equation}
where $Y_{it}$ is gross output, $X^1_{it}, \ldots, X^N_{it}$ are inputs, and $A_{it}$ denotes productivity. 

The firm seeks to minimize the cost of producing any value of gross output $\bar{Y}_{it}$. An input is fully flexible if it is chosen in the current period and affects only current-period costs. Let $\mathcal{V}$ denote the set of fully flexible
inputs. All other inputs are taken as predetermined. The firm takes input prices $W^j_{it}$ as given. The cost-minimization problem is:
\begin{equation}\label{e:cost_min}
\begin{split}
\min_{\{X^j_{it}\}_{j \in \mathcal{V}}} \quad 
    & \sum_{j=1}^N W^j_{it} X^j_{it} \\
\textrm{s.t.} \quad 
    & \bar{Y}_{it} = F(X^1_{it}, \ldots, X^N_{it}, A_{it}).
\end{split}
\end{equation}
The cost-minimizing first-order condition (FOC) for optimal input demand 
is:
\begin{equation}\label{e:FOC}
W^j_{it} = \lambda_{it} \frac{\partial F}{\partial X^j_{it}},
\end{equation}
where the Lagrange multiplier $\lambda_{it}$ is the firm's marginal cost: the shadow value of relaxing the production constraint by one unit. The same marginal cost $\lambda_{it}$ enters the FOC for any input the firm can freely adjust. To interpret this condition, we define three terms:
    \begin{itemize}
     \item $\mu_{it} \equiv \frac{P_{it}}{\lambda_{it}}$ is the firm's markup of output price $P_{it}$ over marginal cost,
     \item $\gamma^j_{it} \equiv \frac{\partial F}{\partial X^j_{it}} \frac{X^j_{it}}{Y_{it}}$ is the elasticity of output with respect to input $X^j_{it}$, and
     \item $s^j_{it} \equiv \frac{W^j_{it}X^j_{it}}{P_{it} Y_{it}}$ is the factor's share in revenue.
    \end{itemize}
With these definitions, the FOC implies the key relationship at the heart of the production approach:
\begin{equation}\label{e:FOC-ratio}
\gamma^j_{it} = \mu_{it}\, s^j_{it}.
\end{equation}
For any input, market power ($\mu_{it}>1$) implies that the output elasticity $\gamma^j_{it}$ exceeds its revenue share $s^j_{it}$. The markup is a property of the firm, so the same $\mu_{it}$ enters the FOC for every input. Hence, any flexible input can, in principle, be used to identify the markup.\footnote{Section \ref{s:right_FOC} discusses what happens when inputs face frictions beyond the static cost minimization problem \eqref{e:cost_min}. For instance, some inputs such as capital may face adjustment costs; others may be subject to input-market power \`{a} la \citet{Robinson1933}. In such cases, the FOC acquires additional terms relative to \eqref{e:FOC-ratio} because the input price $W^j_{it}$ is not taken as given. In principle, one can still estimate markups from such inputs by modeling and measuring the additional frictions.}

Throughout, we typically work with a special case of \eqref{e:production_function_general} that separates inputs into capital $K_{it}$, labor $L_{it}$, and intermediate inputs $M_{it}$:
\begin{equation}\label{e:production_function}
Y_{it} = F(K_{it}, L_{it}, M_{it}, A_{it}).
\end{equation}
For a multi-product firm, $Y_{it}$ is an index over outputs; each input is an index over heterogeneous types. A flexible input $X_{it} \in \{K_{it}, L_{it}, M_{it}\}$ satisfies FOC \eqref{e:FOC-ratio}. We next trace out how Hall and others implemented this approach using data in growth rates. We then turn to its more recent implementation in microdata, which focuses on a single input and, where necessary, estimates production relationships in levels.

\subsection{The production approach in growth rates}
\label{s:growth_rates}
In a series of papers, \citet{HallBPEA1986, Hall1988relation, HallR90} pioneered the production approach by applying FOC \eqref{e:FOC-ratio} to all inputs simultaneously. His motivation was that when output rises in a boom for reasons beyond a firm's own productivity, the gap between output growth and input growth reveals the markup. As \citet[p.~285]{HallBPEA1986} writes, 
\begin{adjustwidth}{1.5em}{0.8em}
\begin{singlespace}
Macroeconomic fluctuations reveal a good deal about market structure. Students of industrial organization have not generally exploited cyclical movements in their research; they have concentrated almost entirely on cross-sectional analysis.
\end{singlespace}
\end{adjustwidth}

In implementing this idea, Hall followed \citet{solow1957technical} and log-differentiated the production function \eqref{e:production_function}. Substituting the output elasticities $\gamma^j_{it}$ in the resulting expression for the revenue shares $s^j_{it}$ and the markup $\mu_{it}$ using FOC \eqref{e:FOC-ratio} yields the ``Hall equation'':
\begin{equation}\label{e:Hall_eq}
\Delta y_{it} = \mu_{it}\!\left(s^K_{it}\Delta k_{it} + s^L_{it}\Delta 
l_{it} + s^M_{it}\Delta m_{it}\right) + \Delta a_{it} \;\equiv\; 
\mu_{it}\,\Delta x_{it} + \Delta a_{it},
\end{equation}
where lowercase letters denote logs, $\Delta x^j_{it} \equiv \Delta \log X^j_{it}$, and $\Delta x_{it}$ is revenue-share-weighted input growth.\footnote{\cite{HallR90}, equation 5.29. Factor-specific technology growth is subsumed into $\Delta a_{it}$ by normalizing the elasticity of output with respect to technology, $\partial \log F / \partial \log A_{it}$, to one. Section~\ref{s:flexible_modeling} discusses non-factor-neutral technology. Classic growth accounting \citep{solow1957technical, JorgensonGriliches1967} imposes $\mu_{it} = 1$. If $\mu_{it} > 1$, the Solow residual equals $(\mu_{it}-1)\Delta x_{it} + \Delta a_{it}$ and varies systematically with input growth.} Appropriately estimated, the relationship between input and output growth reveals the markup $\mu_{it}$. Because the Hall equation applies FOC \eqref{e:FOC-ratio} to all inputs, the approach must grapple with inputs such as capital that may not be fully flexible. \citet{BasuFernald2001_Procyclical} and \citet{BFS2001} generalize the cost-minimization problem \eqref{e:cost_min} to account for dynamic inputs and survey the literature that estimates variants of the Hall equation with industry data. Estimation typically assumes a time-invariant markup.

Most estimates using industry data find gross-output markups on the order of 3 to 6 percent \citep[for example,][]{basu1997returns}. \citet{AhmedFernaldKhan_RTS} update 
the growth-rate approach with industry data from 1949 through 2016 and find markups remain modest. These results are at odds with many recent firm-level findings. Some differences could arise from using industry rather than firm data. However, \cite{deloecker2012markups} use firm-level data and also find relatively modest markups using the Hall growth-rate approach; they find larger ones using their preferred implementation method, which is discussed in Section \ref{s:new_literature} and the remainder of the paper.  

In addition to estimates of markups, the Hall literature provides three lessons that will return regularly in the remainder of this paper.

First, one wants robustness to the functional form of the production function \eqref{e:production_function_general} or \eqref{e:production_function}. Because \eqref{e:Hall_eq} is derived by log-differentiating the production function, in principle it provides a local approximation valid for any production function. There is no need to choose a specific functional form. The functional-form robustness is slightly attenuated by the econometric need to have a time-invariant markup parameter, so it is perhaps best considered a first-order approximation to any production function.

Second, markups are not the same as pure economic profits. For example, a firm might charge a markup to cover fixed or overhead costs. The degree of returns to scale (RTS), $\gamma_{it}$, is the ratio of average cost to marginal cost: 
$\gamma_{it} = \text{AC}_{it}/\text{MC}_{it}$ (Varian 1984, p.~68; \citealp{Syverson2025_ARE}). The firm earns pure economic profits $\Pi_{it}$ if revenues ($P_i Y_i$) exceed the cost of production ($\text{Cost}_{it}=\sum_j W^j_{it}X^j_{it}$). Let the rate of pure economic profit in revenue be $s_{\Pi_{it}}$. Using insights from \cite{HallR90}, \citet{basu1997returns} derive a useful relationship between markups, returns to scale, and the profit rate: 
\begin{equation}\label{e:markups_RTS}
    \frac{\gamma_{it}}{\mu_{it}} = 1 - s_{\Pi_{it}}. 
\end{equation}

The ratio of returns to scale to the markup equals one minus the profit share.\footnote{Section~\ref{s:right_FOC} discusses refinements in the presence of adjustment costs or input-market power such as monopsony. The equation holds with adjustment costs as long as dynamic factors such as capital are priced at their appropriate shadow costs. Monopsony power requires an extra term, since profits can arise from input-price markdowns as well as from output-market markups.} Economic profits arise when markups exceed the degree of returns-to-scale, that is, when $\mu_{it} > \gamma_{it}$. For example, increasing returns ($\gamma_{it} > 1$) could arise from fixed or overhead costs that raise average but not marginal costs. The firm can charge a markup simply to cover those fixed or overhead costs, with no pure profit left over. This distinction matters for interpreting production-based markup estimates: high measured markups need not imply high profits, and rising profit shares need not imply rising markups. 

Third, because of equation \eqref{e:markups_RTS}, the production-approach literature following \cite{HallR90} gave equal weight to markups and returns to scale. Macro modeling that assumed imperfect competition also typically assumed increasing returns in the form of fixed costs to keep profit rates from being too large \citep{RotembergWoodford1999}. The newer literature, by contrast, typically estimates markups while treating returns to scale and output elasticities as nuisance parameters. 

All three lessons remain relevant as we interpret the new literature.

Most of the literature following the Hall growth-rate approach uses industry data. A few papers, however, use firm microdata. These include \cite{Klette1999} and \cite{Dobbelaere_Mairesse2013}. \cite{Klette1999} uses a panel of Norwegian establishments. Rather than taking growth rates over time, Klette takes a local approximation in the cross-section in order to flexibly approximate any functional form. 
\subsection{The new literature on production-based markups}\label{s:new_literature}

The production approach to markup estimation has expanded rapidly with the increasing availability of firm microdata. The new literature typically implements Hall's insights about production-based markups differently than Hall and the literature discussed in the previous subsection. The new literature uses firm or establishment data, focuses on a single fully flexible input, and (typically) estimates production functions in levels. This review focuses on one important modern implementation of the production approach: the firm-level, single-input, levels-based framework that builds primarily on \citet{deloecker2012markups}.


\cite{deloecker2011recovering} and \cite{deloecker2012markups} argue that the growth-rate approach is poorly suited to microdata: first-differencing exacerbates measurement error in short panels \citep{griliches1998production}; in addition, the types of aggregate demand instruments used in the growth-rate literature have low power with limited firm-level cross-sectional variation. (Of course, as Section~\ref{s:econometrics} discusses, not all econometric methods to address endogeneity concerns use instrumental variables.) In addition, the Hall approach assumes that markups are constant over time; the variants used by \cite{Klette1999} and \cite{Dobbelaere_Mairesse2013} impose that the markups are the same for all firms in an industry. \cite{deloecker2012markups} emphasize that they are particularly focused on estimating markups at a firm-by-firm level.

\citet{deloecker2012markups} outline how focusing on a single input---rather than a weighted bundle of inputs as in the Hall approach---can help overcome these shortcomings. The approach has three steps:
\begin{itemize}
    \item correctly select a flexible input $X^j_{it}$,
    \item measure its revenue share $\widehat s^j_{it}$, and 
    \item specify a production function and estimate the output elasticity $\widehat\gamma^j_{it}$.
\end{itemize}
The markup then follows directly from FOC \eqref{e:FOC-ratio}: $\widehat\mu_{it} = \widehat\gamma^j_{it}/\widehat s^j_{it}$.

The first two steps are linked. We need a fully flexible input so that the input's revenue share can potentially be observed in the data at the relevant market price. For quasi-fixed inputs such as capital---chosen in a prior period or subject to adjustment costs---the revenue share depends on an unobservable shadow price, making it more difficult to recover the allocative cost the FOC requires.

For the third step, estimating the production function generally requires including all inputs to avoid omitted-variable bias, even though only one input's elasticity is needed for the markup. Consider a Cobb-Douglas production function in capital, labor, and materials:
\begin{equation}\label{e:pf}
y_{it} = \gamma^K k_{it} + \gamma^L l_{it} + \gamma^M m_{it} + a_{it}.
\end{equation}
For example, omitting capital---even if it is not the input used to measure the markup---leads to omitted-variable bias. Firms that use more capital also tend to use more labor and materials, so the omission will load on their coefficients.

An important question in the third step is how flexibly to model the production technology. The choice of functional form determines how many degrees of freedom output elasticities have to explain revenue shares. The Cobb-Douglas function \eqref{e:pf} is the simplest choice, but it restricts output elasticities to be common across firms and time. Researchers can relax this by allowing elasticities to vary across time or industries, or by moving to a more flexible functional form altogether. A translog production function adds squared and cross terms in the inputs, giving it more flexibility: as a second-order approximation to any arbitrary production technology, it lets elasticities vary with input intensity.\footnote{\label{foot:translog}With two inputs, the baseline translog with Hicks-neutral technology is $y_{it} = \alpha_0 + \alpha_K k_{it} + \alpha_L l_{it} + \frac{1}{2} \beta_{KK} (k_{it})^2 + \beta_{KL} k_{it} l_{it} + \frac{1}{2} \beta_{LL} (l_{it})^2 + a_{it}$. The output elasticity for labor is then $\gamma^L_{it} = \alpha_L + \beta_{KL} k_{it} + \beta_{LL} l_{it}$, which varies across firms with their input intensity.} 

Researchers can also relax assumptions on productivity. One such generalization allows for input-augmenting technical change, where productivity has both a Hicks-neutral component $A^H_{it}$ and, say, a labor-augmenting component $A^L_{it}$. The production function then becomes $Y_{it} = A^H_{it}F(K_{it}, A^L_{it}L_{it}, M_{it})$, which in constant elasticity of substitution (CES) form is:
\begin{equation}\label{e:CES_labor_augmenting}
Y_{it}=A^H_{it} 
\left[ 
(1 - \alpha_l - \alpha_m) K_{it}^{\frac{\sigma - 1}{\sigma}} 
+ \alpha_l \left( A^L_{it} L_{it} \right)^{\frac{\sigma - 1}{\sigma}} 
+ \alpha_m M_{it}^{\frac{\sigma - 1}{\sigma}} 
\right]^{\frac{\sigma}{\sigma - 1}}.
\end{equation}
With this technology, output elasticities depend on both input intensity and the input-augmenting productivity. We return to this choice and its consequences in the next sections.

The production approach is broadly applicable precisely because it rests on cost minimization rather than a specific model of demand or competition. Demand-side approaches, such as \citet{berry1995automobile}, require detailed product-level data, an explicit model of consumer demand, and assumptions about strategic interaction and firm conduct. Such approaches can be highly informative within a narrowly defined industry but require extensive customization for each setting. Cost minimization, by contrast, is a more general principle: regardless of how complex or dynamic the profit-maximization problem may be, firms typically aim to produce at the lowest possible cost. This breadth makes the production approach attractive for cross-industry and aggregate analysis. It comes, however, at a cost.

The production-based markup is measured as a residual: it takes whatever value makes FOC \eqref{e:FOC-ratio} hold given the observed revenue share and the estimated elasticity. This framing has an important implication. Any misspecification---of the FOC itself, of the production function, or of the econometric method used to estimate it---gets absorbed into the estimated markup. So does any mismeasurement of the revenue share. The markup is, in this sense, the residual claimant of the production approach: it ensures the identity holds, absorbing whatever the estimation does not pin down. This is both the approach's strength---it requires minimal assumptions---and its central vulnerability, which the rest of this review unpacks.

\subsection{Key conceptual choices in implementation}\label{s:implementation}

In principle, the method of implementing the production approach laid out in Section~\ref{s:new_literature} is simple: choose a flexible input, measure its revenue share, estimate its output elasticity, and compute the ratio. In practice, each step involves choices that can substantially affect the estimated markups. Which choices create the most sensitivity? We begin with three conceptual issues.

\textbf{Which input to use.} FOC (4) implies any flexible input should yield the same markup. In practice, markups estimated from different inputs often diverge sharply---in both levels and trends. Section \ref{s:garden} documents these divergences using direct comparisons across inputs within the same datasets. Section \ref{s:right_FOC} explains why they might arise.

\textbf{How flexibly to model output elasticities.} Having chosen an input, the researcher must specify a production function and estimate its output elasticity. A Cobb-Douglas function restricts elasticities to be common across firms; a translog or CES function allows them to vary with input intensity. When the assumed functional form is too restrictive, genuine variation in technology gets forced into the markup residual---making measured markups rise or fall even when true market power is unchanged. Section \ref{s:garden} documents how sensitive markup trends are to this choice. Section~\ref{s:flexible_modeling} discusses the resulting bias.

\textbf{Value added versus gross output.} Macroeconomists often want a ``value-added markup'' because GDP aggregates value added across producers. But value-added and gross-output markups are conceptually distinct and cannot be used interchangeably in calibrating macro models. A separate issue is estimation. \citet{BasuFernald1995} show that markups invalidate value added as a production measure except under the Leontief assumption that intermediates enter in fixed proportion to gross output---an assumption with little empirical support. Section~\ref{s:va_vs_go} takes up both issues in detail.

Despite these choices---and the sensitivity to them documented throughout this review---the production approach remains the most broadly applicable tool available for measuring firm-level markups across industries and countries. That is precisely why getting the choices right, and being transparent about them, matters so much.

\section{The garden of forking paths}\label{s:garden}
Existing empirical evidence suggests that the main implementation choices have stark implications for the measurement of markups. What seems like a clean, mechanical implementation quickly gives way to a garden of forking paths.

This section highlights two important forks in that garden: (1) which input to use, and (2) how flexibly to specify technology. These are not the only forks, but they are particularly well documented: Each has been studied empirically using variants of the same data, and each has been shown to materially affect the estimated markups. Later sections explore additional forks, including the specification of the first-order condition (Section~\ref{s:conceptual}), data construction (Section~\ref{s:data}), and econometric method (Section~\ref{s:econometrics}).

We present empirical evidence on these two forks, using studies that offer direct comparisons across choices within the same dataset. We defer most interpretation to Section~\ref{s:conceptual}. A recurring finding is uncertainty about markup levels and trends: it is easy to find plausible specifications in which markups are relatively constant.

\subsection{First fork: Which input to use?}
Any fully flexible input that satisfies the conditions in the cost-minimization problem \eqref{e:cost_min} should yield the same markup. In practice, markup estimates based on different inputs often diverge in both levels and trends; and, the divergences are far too large to dismiss as sampling error. We highlight this fork because, in many applications of the production approach, the input choice is unexamined: researchers simply assert that a given input is flexible.

Table~\ref{t:garden1} presents results from three papers that compare markup estimates across inputs using the same dataset. The differences are stark, and they lead to opposite conclusions about the direction and magnitude of changes in market power.

\begin{table}[H]
\renewcommand{\thetable}{1a}
\centering
\caption{Garden of Forking Paths}\label{t:garden1}
\scriptsize\singlespacing
\renewcommand{\arraystretch}{0.84}
\begin{tabularx}{\linewidth}{p{1.5in} C C}
\multicolumn{3}{c}{\textbf{First Fork: which input to use for the production approach?}} \\
\midrule
\addlinespace[2pt]
& \multicolumn{2}{c}{Production-Based Markups using Different Inputs}\\
\addlinespace[2pt]
\multicolumn{1}{l}{\textbf{Raval (2023)}} & Labor & Materials \\
 \cmidrule(r){2-2} \cmidrule(r){3-3}
U.S. (1970--2010) & Increase of 90\% & Decrease of 50\%  \\
Chile (1978--1996) & Decrease of 20\% & Increase of 15\%  \\
Colombia (1978--1996) & Decrease of 30\% & Increase of 10\%  \\
India (1998--2014) & Decrease of 40\% & No change \\
Indonesia (1991--2000) & Decrease of 10\% & Increase of 5\% \\
\addlinespace[2pt]
\multicolumn{2}{l}{\textbf{Doraszelski \& Jaumandreu (2019)}}\\
& Labor & Materials \\
\cmidrule(r){2-2} \cmidrule(r){3-3}
Spain (1990-2012) & Exporters charge higher markups & Exporters charge smaller markups \\
\addlinespace[2pt]
\multicolumn{1}{l}{\textbf{Raval (2023)}} & Energy & Non-Energy (Raw) Materials \\
  \cmidrule(r){2-2} \cmidrule(r){3-3}
Chile (1978--1996) & Increase of 20\% & Increase of 15\% \\
Colombia (1978--1996) & Decrease of 70\% & Increase of 20\% \\
India (1998--2014) & Decrease of 25\% & Increase of 5\% \\
Indonesia (1991--2000) & Increase of 40\% & Decrease of 5\% \\
\addlinespace[2pt]
  \multicolumn{1}{l}{\textbf{Traina (2018)}}& Cost of Goods Sold & Operating Expenses \\
  \cmidrule(r){2-2} \cmidrule(r){3-3}
Compustat, 1950--2016 & Increase: 1.19 to 1.45 & Increase: 1.15 to 1.17 \\
\addlinespace[1pt]
\hline
\multicolumn{3}{p{\dimexpr\linewidth-2\tabcolsep}}{{Sources: Figures 2 and 5 from Raval (2023), Table 1 from Doraszelski \& Jaumandreu (2019), and Figure 2 from Traina (2018). Raval and Doraszelski \& Jaumandreu entries report approximate percentage changes in average markups, rounded to nearest 5 percent. Traina entries report markup levels.}}
\end{tabularx}
\renewcommand{\arraystretch}{1.0}
\end{table}

The input choice can reverse the sign of markup trends entirely. \citet{raval2023testing} uses manufacturing census data from five countries to show labor- and materials-based markup estimates diverge far beyond what sampling error could explain, implying at least one set of identifying assumptions is wrong. In the United States, labor-based markups nearly double between 1970 and 2010, while materials-based markups fall by half. In Chile and Colombia, labor-based markups fall while materials-based markups rise. Within all five countries, labor and materials markups are negatively correlated across firms, in both levels and trends. These patterns persist across specifications and estimation methods.

The input choice also shapes cross-sectional comparisons, not just trends. \citet{doraszelski2019using} compare exporters and non-exporters in Spain and find the conclusion about which group charges higher markups flips depending on the input used. Exporters appear to charge higher markups when labor is used, but lower markups when materials are used instead.

Even within a single broad input category, the choice of subset matters. The middle panel of Table~\ref{t:garden1} shows additional results from \citet{raval2023testing} comparing energy and non-energy materials. In Colombia, markups based on energy fall 70\% while markups based on non-energy materials rise 20\%. In India and Indonesia, the markup trends switch signs entirely. These patterns matter because input flexibility is often assumed rather than demonstrated: implementations of the production approach frequently use materials on the asserted grounds that they are more flexible than labor, yet even within materials, markup estimates diverge sharply across subsets.

The same sensitivity appears in financial data. Studies using Compustat---the most widely used dataset for broad-coverage markup estimation---must choose between Cost of Goods Sold (COGS), which covers direct production costs, and Operating Expenses (OPEX), which adds Selling, General, and Administrative expenses (SG\&A) such as marketing, executive salaries, and R\&D. \citet{traina2018aggregate} finds this choice is decisive. Using COGS alone implies markups rose from 1.19 to 1.45 between 1950 and 2016, a 26 percentage point increase. Using OPEX, which includes SG\&A, implies markups rose from 1.15 to 1.17, an increase of just 2 percentage points. Including SG\&A wipes out most of the rise. This sensitivity has both a measurement and an econometric interpretation. From a measurement perspective, SG\&A may simply be a variable cost that belongs in the revenue-share denominator. From an econometric perspective, even if SG\&A is quasi-fixed, omitting it from the production function biases the estimated elasticity of the included input upward, just as omitting capital does. We develop these interpretations in Section~\ref{s:data}.

\citet{BenkardMillerYurukoglu2025} document a related concern: \citet{deloecker2020rise}'s replication code drops all observations with missing SG\&A---27\% of the sample---even though SG\&A is not used in their markup calculation. Firms with missing SG\&A systematically have higher COGS ratios and lower measured markups, creating trended upward selection bias. Both findings implicate SG\&A in the measured markup increase, though through different mechanisms: input measurement in \citet{traina2018aggregate} and sample construction in \citet{BenkardMillerYurukoglu2025}. The interpretation remains contested. \citet{DeLoeckerEeckhout2025reply} reply and defend their original sample selection; \citet{BenkardMillerYurukoglu2025b} reply to the reply.

The choice of input shapes substantive conclusions about markups. While the logic of the production approach implies any flexible input should yield the same markup, different inputs---whether labor, materials, energy, or accounting aggregates---often imply divergent and even contradictory trends. Nearly every input is subject to some friction that can drive a wedge between the observed price and the allocative shadow price the FOC requires. If frictionless input markets are the exception rather than the rule, the divergence across inputs in Table~\ref{t:garden1} is what we might naturally expect.

\subsection{Second fork: How flexibly to model output elasticities?}\label{s:flexibility_fork}
Having chosen a flexible input, the researcher must decide how to estimate its output elasticity. This choice determines how many degrees of freedom output elasticities have to explain variation in revenue shares, and therefore how much of that variation gets attributed to technology versus to the markup residual. When the assumed functional form is too restrictive, genuine variation in output elasticities gets forced into estimated markups, making them appear to rise or fall even when true market power is unchanged.

Table~\ref{t:garden2} documents the consequences of different elasticity estimation strategies in three studies that apply multiple approaches to the same data.

\begin{table}[H]
\renewcommand{\thetable}{1b}
\centering
\caption{Garden of Forking Paths}\label{t:garden2}
\scriptsize\singlespacing
\renewcommand{\arraystretch}{0.84}
\begin{tabularx}{\linewidth}{p{1.1in} C C C}
\multicolumn{4}{c}{\textbf{Second Fork: how to estimate the output elasticity for a given input?}} \\
\midrule
\multicolumn{3}{l}{\textbf{De Loecker, Eeckhout \& Unger (2020)}}\\
\multicolumn{3}{l}{Compustat, 1980-2016}\\
 & Cost Share & Cobb-Douglas & Overhead \\
 \cmidrule(r){2-2} \cmidrule(r){3-3} \cmidrule(r){4-4}
 & Increase: 1.29 to 1.73 & Increase: 1.21 to 1.61 & Increase: 0.98 to 1.32 \\
\addlinespace[2pt]
\multicolumn{3}{l}{\textbf{Foster, Haltiwanger \& Tuttle (2024)}}\\
\multicolumn{3}{l}{U.S. Manuf., 1977-2012}\\
 & Cost Share & Cobb-Douglas & Translog \\
 \cmidrule(r){2-2} \cmidrule(r){3-3} \cmidrule(r){4-4}
4-digit industry & Increase: 1.25 to 1.50 & Stable: 1.25 to 1.25 & Decrease: 1.00 to 0.99\\
2-digit industry & Increase: 1.40 to 1.80 & Increase: 1.35 to 1.48 & Increase: 1.30 to 1.60 \\
\addlinespace[2pt]
  \multicolumn{1}{l}{\textbf{Demirer (2025)}} & & Cobb-Douglas & Labor-augmenting \\
 \cmidrule(r){3-3} \cmidrule(r){4-4}
US (1961--2018) & & Increase: 1.30 to 1.50 & Increase: 1.25 to 1.30 \\
\multicolumn{2}{l}{Chile (1979--1996)} & Increase: 1.35 to 1.40 & Decrease: 1.26 to 1.22\\
\multicolumn{2}{l}{Colombia (1978--1991)}  & Stable: 1.40 to 1.40 & Decrease: 1.30 to 1.28\\
\multicolumn{2}{l}{India  (1998--2014)} & Stable: 1.31 to 1.31 & Increase: 1.20 to 1.29\\
\multicolumn{2}{l}{Turkey (1983--2000)} & Increase: 1.25 to 1.31 & Decrease: 1.20 to 1.10\\
\addlinespace[1pt]
\hline
\multicolumn{4}{p{\dimexpr\linewidth-2\tabcolsep}}{{Sources: Figures 7A, 1 and 8A from De Loecker, Eeckhout \& Unger (2020), Figures 2 and 3 from Foster, Haltiwanger \& Tuttle (2024), and Figures 6 and OA-5 from Demirer (2025). Entries report the direction and approximate markup levels at the start and end of each sample period. De Loecker, Eeckhout \& Unger (2020) use cost of goods sold as the flexible input. Foster, Haltiwanger \& Tuttle (2024) use materials. Demirer (2025) uses the combination of labor and materials.}}
\end{tabularx}
\renewcommand{\arraystretch}{1.0}
\end{table}

One way to forego more elaborate estimation and specification of 
technology is to proxy output elasticities with shares of payments to each factor in total costs. As Section \ref{s:flexible_modeling} discusses, under the 
assumption of constant RTS and competitive input markets, these cost shares 
equal output elasticities; no additional estimation required. 
\citet{deloecker2020rise} implement a version of this by fixing the 
output elasticity at its historical average cost share across firms, 
applying the same value to every firm in a given industry and period. 
They find this yields markup trends broadly similar to their 
Cobb-Douglas benchmark---markups rising from around 1.29 to 1.73 under
cost shares, compared to 1.21 to 1.61 under Cobb-Douglas---suggesting 
the upward trend is not merely an artifact of the estimated elasticity 
rising over time. Even when they extend the production function to include a fixed (overhead) input, the upward trend in markups remains.\footnote{This specification, which they call PF2, includes SG\&A as a productive input along with COGS and capital. It is similar to the \citet{traina2018aggregate} OPEX specification from Table \ref{t:garden1}, except that they do not combine COGS and SG\&A. Unlike Traina, they still find a sharply rising markup. The PF2 specification and the associated profit analysis were added in the published version; the earlier working paper used COGS as the sole variable input without discussing overhead as a potential production input.}

That similarity between their cost-share and Cobb-Douglas estimates is revealing: the estimated output elasticity is not doing much independent work across the two specifications. But it also reflects the fact that using an average cost share gives output elasticities essentially no additional degrees of freedom to explain cross-firm variation in revenue shares relative to Cobb Douglas. An alternative, which we discuss in Section~\ref{s:flexible_modeling}, is to estimate a separate output elasticity for each firm and period; this alternative gives elasticities more room to explain the data before the markup residual absorbs what remains. Cost-share proxies carry their own assumptions, and the choice between them and more elaborate elasticity estimates is itself a fork in the garden.

Even within a single functional form, the level of industry aggregation at which elasticities are estimated can reverse markup trends entirely. \citet{FosterHaltiwangerTuttle2024} use U.S. Census of Manufacturing plant data from 1977 to 2012, with materials as the flexible input. Under Cobb-Douglas at the 2-digit industry level, markups rise from 1.35 to 1.48. At the 4-digit level, they are stable at 1.25 throughout.\footnote{Industry-level aggregates such as KLEMS data do exist but average over the firm heterogeneity that drives markup dispersion; the production approach's value is precisely that it recovers firm-level markups. The \citet{FosterHaltiwangerTuttle2024} results underscore the point: 2-digit and 4-digit industry aggregation yield different markup trends from identical plant-level data.} The mechanism is simple: coarser industry groupings restrict how much output elasticities can vary across firms, forcing more cross-firm technological heterogeneity into the residual claimant of FOC \eqref{e:FOC-ratio}, the markup.

Moving to a more flexible functional form changes the picture further, and its effect interacts with the aggregation level. Under translog at the 4-digit level, \citet{FosterHaltiwangerTuttle2024} find markups decline slightly (1.00 to 0.99)---where Cobb-Douglas showed stability and cost shares showed a rise. At the 2-digit level, translog produces a larger rise (1.30 to 1.60) than Cobb-Douglas. The underlying plant-level revenue shares are identical across all specifications: any divergence reflects how different estimation choices assign variation between output elasticities and markups through FOC \eqref{e:FOC-ratio}, not differences in the data.

Assumptions about the nature of productivity matter as much as functional form. \citet{Demirer2025} finds that moving beyond Hicks-neutral productivity to incorporate labor-augmenting productivity reverses markup trends in multiple countries. In U.S. manufacturing, Cobb-Douglas with Hicks-neutral technology implies markups rose from 1.30 to 1.50 between 1961 and 2018; labor-augmenting CES implies a much more modest increase from 1.25 to 1.30. In Chile and Turkey the trends flip sign: rising under Cobb-Douglas but falling under labor-augmenting CES. The rationale, developed further in Section~\ref{s:flexible_modeling}, is that labor-augmenting productivity gives output elasticities an additional degree of freedom to explain changes in relative factor costs, leaving less variation in input shares $\widehat s^j_{it}$ to load onto the markup. 

In sum, the choice of technology specification shapes substantive conclusions about markups just as much as the choice of input does. The logic of the production approach implies that output elasticities should absorb genuine technological variation. But restrictive functional forms, industry aggregation, or assumptions about the nature of technological progress can all limit how much variation in revenue shares gets attributed to technology rather than to the markup. When the specification restricts how output elasticities can vary, the variation that elasticities cannot absorb gets loaded onto the markup instead.

\section{Conceptual rationales for the garden} \label{s:conceptual}
This section highlights two theoretical hypotheses that can help reconcile the discrepancies from the garden of forking paths: (1) the central first-order condition could be missing non-markup frictions and (2) the estimated production technology could be insufficiently flexible. These two hypotheses can, in principle, generate the observed empirical divergences, and they both have support in the literature. They are not the only hypotheses, since there could also be non-conceptual sources of bias arising from inadequacies in the data (Section \ref{s:data}) or econometric issues in estimating output elasticities (Section \ref{s:econometrics}).

\subsection{Do we have the right first-order condition?}\label{s:right_FOC}
The key FOC \eqref{e:FOC-ratio} assumes a static cost-minimization problem in which the firm takes input prices as given. When these assumptions fail, the effective cost of an input to the firm, that is, the price that actually governs its hiring decision, differs from the observed market price. This gap introduces an unobserved wedge $\tau^j_{it}$ between the output elasticity $\gamma^j_{it}$ and the revenue share $s^j_{it}$:\footnote{We follow \citet{doraszelski2019using} in writing all deviations from FOC \eqref{e:FOC-ratio} in this form.}
\begin{equation} \label{e:FOC_ratio_generic_wedge}
\gamma^j_{it} = \mu_{it} s^j_{it} \left( 1+\tau^j_{it} \right) \text{.}
\end{equation}

Suppose we ignore the wedge. Even with correct values for $\gamma^j_{it}$ and $s^j_{it}$, the production approach recovers $\mu_{it}(1+\tau^j_{it})$, not the true markup $\mu_{it}$. Suppose also that different inputs face different wedges. They will then yield different markups---the pattern in Table~\ref{t:garden1}. 

A fundamental implication of \eqref{e:FOC_ratio_generic_wedge} is that a single FOC cannot separately identify two unknowns. Observed revenue shares $s^j_{it}$ reflect both the true markup $\mu_{it}$ and the wedge $\tau^j_{it}$, but the FOC alone cannot tell us how much revenue-share variation is due to one versus the other. Since heterogeneous markups are themselves a form of distortion, they could in principle be folded into the broader category of wedges. But not all wedges are markups, and conflating the two leads to misinterpretation. Separately identifying markups from other wedges determines whether we are measuring market power or something else entirely.

Separate identification requires bringing in additional modeling assumptions, and typically additional data. The choice of what to assume determines what gets identified. In quantifying misallocation, \citet{restuccia2008policy} and \citet{hsieh2009misallocation} impose structure on markups---assuming common output elasticities $\gamma^j_t$ within industries and a common demand elasticity that pins down the markup, so that all firm-specific variation in revenue shares is absorbed by exogenous distortion wedges $(1+\tau^j_{it})$.\footnote{\citet{restuccia_rogerson2017} survey the misallocation literature; see also \citet{HKS:2018} on measurement. \citet{peters2020heterogeneous, edmond2023how, baqaee2020productivity, Baqaee2024} discuss heterogeneous markups as a source of misallocation.} \citet{deloecker2012markups} take the opposite approach: they impose structure on the wedges (by effectively assuming they are common across firms), which allows all variation in revenue shares to be attributed to markups. Both approaches achieve identification, but by assumption rather than by data. A third path is to render either markups or wedges, or both, a function of observables. \citet{CLRS:2024}, for instance, use a demand model to quantify markups and then use the residual variation in revenue shares to identify input-market wedges.

These wedges can arise for three distinct reasons, each discussed in turn below.

The first source of wedges arises when observed input prices are not the allocative prices that govern firms' optimization. Even in a static setting, government-induced frictions---tariffs, taxes, quotas, or size-based regulations---can drive a gap between the price a firm pays and the shadow cost it responds to. 
A specific form of this wedge is input-market power \citep{Robinson1933}. Concretely, suppose the firm is a monopsonist in the labor market. It faces upward-sloping labor supply: it must pay a higher wage to hire more. The firm then acts as if the wage were $W^j_{it}(1+\epsilon^j_{it})$, where $\epsilon^j_{it}$ is the elasticity of the input price with respect to quantity hired. The result is a ``markdown'' of the observed wage relative to the marginal revenue product. Production-based markup estimates will conflate that markdown wedge with markup $\mu_{it}$. Following \citet{Dobbelaere_Mairesse2013}, several papers use the divergence of markup estimates across inputs to quantify input-market power, treating the gap between $\hat\mu^L_{it}$ and $\hat\mu^M_{it}$ as informative about the markdown rather than as a nuisance.\footnote{Studies of labor-market power have a long history \citep[such as][]{Manning2003}. In addition to \citet{Dobbelaere_Mairesse2013}, recent examples include \citet{berger2022labormarketpower, YehMacalusoHershbein2022, jarosch2024granular, kirov2023labor, rubens2023market}.} 

The second source of wedges arises when the cost-minimization problem is inherently dynamic, yet the researcher imposes a static framework. Capital is the classic case: today's investment affects future production costs, so the relevant shadow price of capital incorporates adjustment costs and future expectations, not just today's rental rate. As \citet{basu2002aggregate} and \citet{doraszelski2019using} show, the resulting shadow price can still be written as $W^j_{it}(1+\tau^j_{it})$, where $\tau^j_{it}$ depends on stocks, flows, and expectations, so the factor-demand equation retains the form of \eqref{e:FOC_ratio_generic_wedge}.\footnote{\citet[p.~245]{BFS2001} and \citet{doraszelski2019using} consider adjustment costs in the context of production approach to markups. See also \citet{BerndtFuss1986} and \citet{Hulten1986}.} Adjustment costs are not limited to capital, nor are they inherently inefficient \citep{asker2014dynamic}. \citet{CooperHaltiwangerWillis} find labor adjustment costs are important and rising in U.S. manufacturing; ignoring them and treating labor as fully flexible generates what they estimate as substantial and rising dispersion in production-based markups despite there being no variation in actual markups. Even materials and other intermediate inputs may face dynamic frictions through supply-chain relationships and delivery lags \citep{LiuTsyvinski2024, DhyneEtAl2022, AcemogluTahbaz2025}.

A related concern is that observed wages may not equal the allocative shadow cost even if the firm minimizes costs and there are no adjustment costs or input-market frictions. Most employment relationships are long-term. \citet{Hall1980_Employment} argues wages should be understood as installment payments on a firm's long-term obligation to its workers, not as spot prices that reflect the current marginal value of labor. If so, the observed wage at any point in time can differ substantially from the shadow cost that governs the firm's current hiring decision \citep{BasuHouse2016, Kudlyak2024}. Similar issues can arise in firm-to-firm transactions governed by relational contracts, where observed transfer prices may reflect long-term relationship value rather than current marginal cost.\footnote{\citet{Rosen1985} surveys implicit contracts in labor markets. \citet{Macchiavello_Rocco_Relational} survey relational contracts between firms.} In both cases, using observed prices in place of allocative shadow costs introduces a time-varying wedge $\tau^j_{it}$ into \eqref{e:FOC_ratio_generic_wedge}.

The third source of wedges arises when the firm's optimization problem does not take the form of cost minimization at all. Wage bargaining is the leading example, as in search-and-matching models.\footnote{\citet{RogersonShimerWright2005} survey search-and-matching models of the labor market, which are also inherently dynamic. They discuss different specifications of the bargaining process.} The outcome of the bargaining problem can be interpreted as a wedge $\tau^j_{it}$ in equation \eqref{e:FOC_ratio_generic_wedge} \citep{doraszelski2019using}.\footnote{Here is a simple bargaining model where the outcome yields equation \eqref{e:FOC_ratio_generic_wedge}. The wage is set to split the match surplus; the worker's outside option is $U_t$ and the worker's bargaining share is $\beta_t$. The match surplus is the difference between the worker's marginal product and outside option: $W^j_{it} = U_t + \beta_t\!\left(\frac{P_{it}}{\mu_{it}} 
\frac{\partial F_{it}}{\partial X^j_{it}} - U_t\right).$
Rearranging yields an expression of the form 
\eqref{e:FOC_ratio_generic_wedge}, where
\label{e:bargaining_ratio}
$\left(1+\tau^j_{it}\right) \equiv \left[\frac{1}{\beta_t} - 
\frac{1-\beta_t}{\beta_t}\cdot\frac{U_t}{W^j_{it}}\right]$.} For example, even if true markups are constant, changes in bargaining power or workers' outside options can change the implied $\tau^j_{it}$ and the estimated markup $\hat\mu_{it}$.

In short, frictionless input markets are likely the exception, not the rule. The three sources of wedges discussed above---non-allocative prices, dynamic frictions, and non-cost-minimizing behavior---are all empirically relevant and likely widespread. Differences in estimated markups across inputs, as in Table~\ref{t:garden1}, most plausibly reflect differences in these unobserved wedges $\tau^j_{it}$ rather than differences in true markups. That divergence is itself a diagnostic: it signals the presence of frictions or misspecification, not necessarily rising market power. At the same time, the quantitative importance of these wedges for average markup \textit{levels} remains an open question. If adjustment-cost wedges fluctuate around their steady-state value, longer-run markup trends may be more reliable than period-by-period estimates---though even steady-state adjustment costs need not be zero \citep{BFS2001}.

\subsection{How (flexibly) should we model production?}\label{s:flexible_modeling}
A second conceptual hypothesis for the discrepancies in the garden is that the assumed production function is too restrictive. When the production function is too restrictive, variation in revenue shares that truly reflects technological differences across firms will instead load onto the only other object allowed by FOC \eqref{e:FOC-ratio}, the markup. This is the mirror image of the wedge problem in Section~\ref{s:right_FOC}. There, the FOC is misspecified. Here, the FOC may be correctly specified but the production function imposes a wedge $\tau^j_{it}$ between the true output elasticity $\gamma^j_{it}$ and the measured one: $\widehat\gamma^j_{it} = \frac{1}{1+\tau^j_{it}}\gamma^j_{it}$. If a researcher observes $s^j_{it}$ but the estimated elasticity $\widehat\gamma^j_{it}$ is compressed by a restrictive functional form, the production approach recovers $(1+\tau^j_{it})\mu_{it}$ rather than the true markup $\mu_{it}$.

\textbf{How restrictive functional forms load variation onto markups.}
The ratio of FOCs across two inputs shows precisely how a restrictive functional form can generate divergences in measured markups. Suppose labor and materials are both fully flexible and satisfy FOC \eqref{e:FOC-ratio}. Let $\hat{\mu}^j_{it}$ denote the markup estimated using input $X^j$. For each firm $i$, the ratio of the estimated FOCs for labor to materials exactly satisfies:
\begin{equation}\label{e:cost_labor_to_materials}
\frac{s^L_{it}}{s^M_{it}}\cdot
\frac{\hat{\mu}^L_{it}}{\hat{\mu}^M_{it}} 
= \frac{\hat{\gamma}^L_{it}}{\hat{\gamma}^M_{it}}.
\end{equation}
With Cobb-Douglas estimated at the industry level, the right-hand side is constant across firms, so $\hat{\mu}^L_{it} / \hat{\mu}^M_{it}$ must account for all variation in $s^L_{it}/s^M_{it}$. If firms truly share the same Cobb-Douglas function---so the true output elasticities $\gamma^L$ and $\gamma^M$ are identical across firms---then any cross-firm variation in $\hat{\mu}^L_{it} / \hat{\mu}^M_{it}$ reflects only sampling error. If not, the restriction forces technology differences into the markup ratio.

We could permit more variability in output elasticities through two main approaches: estimating at a finer level of aggregation or using a more flexible functional form. Each offers different degrees of freedom for elasticities to explain variation in revenue shares.

One simple fix is to estimate elasticities at finer industry levels. Even with a Cobb-Douglas form, allowing coefficients to vary across four-digit industries rather than pooling at two digits lets elasticities explain more of the observed variation in revenue shares. Coarser estimates risk conflating markup and technology heterogeneity. Table \ref{t:garden2} shows exactly this: at two digits markups rise, at four digits they don't---evidence coarser elasticities force technology variation into measured markups \citep{FosterHaltiwangerTuttle2024}. In practice, however, going much finer is difficult: estimating separate production functions at the firm or product level is practically impossible with the sample sizes available in most datasets.

\citet{doraszelski2019using}, \citet{raval2023testing}, and \citet{Demirer2025} go further, arguing the resolution is to use a more flexible functional form and to relax Hicks-neutrality. Labor-augmenting technical progress can potentially resolve the markup discrepancies in both Tables \ref{t:garden1} and \ref{t:garden2}---even when inputs are fully flexible and do not face non-markup wedges $\tau^j_{it}$. Suppose the production function is CES with labor-augmenting technical progress, as in \eqref{e:CES_labor_augmenting}. Then the ratio \eqref{e:cost_labor_to_materials} becomes:
\begin{equation}\label{e:FOC_relative_CES}
\frac{s^L_{it}}{s^M_{it}} \cdot \frac{\hat{\mu}^L_{it}}{\hat{\mu}^M_{it}}
= \biggl(\frac{\alpha_l}{\alpha_m}\biggr)
  \biggl(\frac{L_{it}}{M_{it}}\biggr)^{\!\frac{\sigma-1}{\sigma}}
  (A^L_{it})^{\frac{\sigma-1}{\sigma}}
\end{equation}
Relative output elasticities $({\gamma^L_{it}}/{\gamma}^M_{it})$ now depend on relative factor intensities $(L_{it}/M_{it})$ and labor-augmenting productivity $A^L_{it}$. Even if $A^L_{it}$ is common across firms within a period (that is, $A^L_{it} = A^L_t$), this extra degree of freedom can explain variation in relative factor costs instead of forcing it into markup differences.
 \citet{Demirer2025} finds allowing for labor-augmenting productivity reduces estimated markup levels and flattens their upward trend compared with Cobb-Douglas or Hicks-neutral CES/translog specifications.\footnote{The labor-augmenting view rests on difficult-to-test assumptions about the nature of productivity. Since at least \citet{solow1957technical} and \citet{sato1967}, it has been recognized technological bias is difficult to distinguish from differential patterns of substitution.}

A more flexible production function can give ${\hat{\gamma}^L_{it}}/{\hat{\gamma}^M_{it}}$ more scope to fit the data. Three questions guide this choice:
\begin{itemize}
\item At what level of aggregation should industries be grouped for estimation?
\item How flexible should the production function be?
\item Is productivity Hicks-neutral, or input-biased (such as labor-augmenting)?
\end{itemize}
Each can matter through the same mechanism: allowing output elasticities to vary more across firms. When elasticities vary, differences in revenue shares can be attributed to technology rather than to markups.

\textbf{Constant returns to scale: a useful restriction with real 
costs.}
Constant RTS is the most common functional form restriction in the production approach, and it has a clear upside: under constant RTS and competitive input markets, output elasticities equal cost shares directly, requiring no additional estimation. To see why, recall from Section~\ref{s:growth_rates} that the degree of returns to scale $\gamma_{it}$ equals the ratio of average to marginal cost, and the markup equals price over marginal cost. Together these imply:
\begin{equation}\label{e:cshare}
\gamma^j_{it} = \gamma_{it}\, c^j_{it},
\end{equation}
where $c^j_{it} = W^j_{it}X^j_{it}/\text{Cost}_{it}$ is input $j$'s share of total costs. When RTS are constant ($\gamma_{it} = 1$) and input markets competitive, the output elasticity equals the cost share directly: $\gamma^j_{it} = c^j_{it}$. This is appealing because cost shares vary across firms and over time, giving output elasticities firm-specific variation that a single industry-level Cobb-Douglas coefficient cannot provide. That additional variation means cost shares can potentially absorb more of the cross-firm heterogeneity in revenue shares, leaving less to be loaded onto the markup. 

The cost of this approach is that it replaces one set of difficult-to-test assumptions with another. Rather than assuming a restrictive production function---whose estimation will also encounter the endogeneity concerns discussed in Section~\ref{s:econometrics}---the researcher assumes RTS are constant and that input markets are competitive. None of these assumptions are easy to verify. Just as a misspecified production function loads true technology variation onto markups, incorrect assumptions regarding RTS and input markets also lead to mismeasurement.

If true returns to scale vary across firms or over time---because of fixed costs, scale economies, or technology differences---that variation would be interpreted as markup variation. The same is true of deviations from perfect competition in input markets: if monopsony power varies across firms and over time, using cost shares will load that variation into the markup. Section~\ref{s:right_FOC} engaged with deviations from perfect competition in input markets. Deviations from constant RTS are equally plausible. \citet{ruzic2021returns} find that a secular decline in returns to scale---from increasing toward constant---helps rationalize long-run trends in factor shares and improves the measurement of misallocation, suggesting that time-varying RTS are empirically relevant and that assuming them away loads their trend into rising markups instead. \citet{BFK2006technology} and \citet{AhmedFernaldKhan_RTS} find evidence consistent with approximately constant RTS at the three-digit industry level, but heterogeneity across industries and firm types may still matter for cross-sectional comparisons. \citet{McAdam2024} find in a panel of European firms that the typical four-digit industry has close to constant returns but that there is considerable heterogeneity. 

\subsection{Gross output versus value added}\label{s:va_vs_go}
A final specification choice---whether to use a gross-output or value-added production function---is conceptually distinct from functional form flexibility but equally consequential. So far, we have assumed that firms produce gross output via production function \eqref{e:production_function}, combining capital and labor with intermediate inputs purchased from other firms.

But many papers in the production-approach literature have chosen to estimate value-added production functions. Value added nets out intermediates and, as a production measure, assumes that firms produce real value added as a function of capital, labor, and technology. Nominal value added is the value of gross output less the cost of intermediate inputs: $P_{it} Y_{it} - W^M_{it} M_{it}$. Real value added in firm-level data is gross output less (real) intermediate inputs: $Y_{it}-M_{it}$.\footnote{\cite{BasuFernald1995} discuss alternative definitions of value added as production measures. In national accounting, real value added is implicitly defined such that a chain-aggregate of materials and value added yields real gross output. } Conceptually, real value added is an unintuitive construct: gross output is shoes, while value added is ``shoes lacking leather, made without power'' \citep[][p.761]{Domar1961}. 

Is value added nevertheless a valid measure of output in terms of estimating output elasticities and markups? Our view is no. The use of value added as a production measure is rejected by the available empirical evidence. \citet{gandhi2017heterogeneous} show that imposing a value-added structure inflates measured productivity dispersion by a factor of five relative to a gross-output specification. Even with perfect competition, value added requires separability: $Y_{it}=F(V(K_{it}, L_{it}, A_{it}), M_{it})$. \cite{Ruzic:2024} finds that this separability assumption fails, echoing earlier findings (such as \citealp{JorgensonGollopFraumeni1987}). Under imperfect competition, the case weakens further: \citet{BasuFernald1995} find, in the context of the Hall-style growth-rate literature (Section \ref{s:growth_rates}), that using value added leads to biased markup estimates. 

Regardless of whether estimation is done with gross output versus value added, macroeconomists often want value-added measures. After all, the broadest measure of aggregate output, GDP, nets out those intermediate transactions by definition. The approach in the representative-firm macro literature has been to take gross-output markups and convert them to equivalents in value-added terms. Just as value added and gross output are different (but related) objects, gross-output markups cannot be simply plugged into a macro model as if they were value-added markups. 

The mapping may be model-specific. \citet{RotembergWoodford1995} derive the most commonly used mapping from gross-output to value-added markups in a representative-firm model where each firm purchases intermediate inputs in fixed proportion to gross output. Because every firm in a supply chain charges a markup, the cumulated value-added markup exceeds the gross-output markup through double marginalization. Specifically, if $s^M$ denotes the intermediate input share of revenue:
\begin{equation}\label{e:VA_GO_markup}
\mu^V = \frac{1 - s^M}{1 - \mu s^M}\cdot\mu.
\end{equation}
Since the gross-output markup $\mu$ exceeds one, the multiplier $(1-s^M)/(1-\mu s^M)$ also exceeds one, so value-added markups are always at least as large as their gross-output counterparts. The gap grows quickly. \citet{deloecker2020rise} find a sales-weighted average gross-output markup of 1.6 in recent years. With an economy-wide intermediate input share of 50\%, equation \eqref{e:VA_GO_markup} would imply a corresponding value-added markup of 4. As \citet{basu2019price} highlights, taken as the relevant macro parameter, a value-added markup of that magnitude would imply strongly negative technological change and a negative return to capital---implications that strain credulity. But the calculation does make clear that the average gross-output markup cannot simply be plugged into a representative-firm macro model.

\subsection{Takeaways from the garden of forking paths}
The garden of forking paths illustrates disparate markup trends need not be mere statistical noise---they can reflect genuine conceptual choices. The forks matter, and researchers face several options for navigating them.

First, the choice of input should be grounded in a persuasive case the central FOC is likely to hold. This requires arguing from institutional detail rather than default conventions. In some contexts, that might mean focusing on a subset of materials---such as energy inputs in industries where they are purchased on competitive spot markets---or on categories of labor, such as temporary-contract workers, whose wages are more likely to reflect contemporaneous market conditions than long-term contracts (such as \citealt{vanHeuvelen_Dual_2021}). Making the case for flexibility is ultimately about convincing the reader the observed input price is the allocative one to which the firm responds when minimizing costs.

Second, when plausible frictions---such as input market power, adjustment costs, or regulatory constraints---threaten the link between the first-order condition and markups, the researcher may need to add more structure to the estimation by modeling the source of the wedge.\footnote{For example, \citet{CLRS:2024} model product-market power, input-market power, and input-specific distortions inside the same FOC, identifying two with model structure and inferring one as a residual.} One can use two inputs to estimate two frictions (such as markups and markdowns); or incorporate dynamic optimization to account for capital or labor adjustment costs; or use auxiliary data to calibrate or instrument for these frictions.\footnote{Examples of each approach include \citet{Dobbelaere_Mairesse2013, doraszelski2019using, kirov2023labor, CooperHaltiwangerWillis}.} Such structure can help avoid loading non-markup distortions into the measured markup.

Third, researchers can choose a production technology that is sufficiently flexible to explain variation in revenue shares without forcing it into the markup residual. This might involve estimating elasticities at a more granular industry level, moving beyond Cobb-Douglas to CES or translog specifications, or allowing for input-augmenting technical change. These choices expand the scope for output elasticities to capture genuine technological heterogeneity so it is not mistaken for market power.

Ultimately, the production approach's central vulnerability---that the markup absorbs all misspecification---is also a guide to best practice. If the estimates from different inputs agree, that convergence is evidence that wedges are small and the production function is well specified. If they disagree, as they often do in Tables~\ref{t:garden1} and~\ref{t:garden2}, that divergence is itself informative: it signals the presence of frictions or misspecification that the researcher should investigate rather than paper over. A complementary approach is to focus on longer-run averages and trends rather than period-by-period estimates. If the frictions that contaminate individual markup estimates are roughly mean-zero over time, longer-run trends may still be informative about market power even when cross-sectional comparisons are unreliable \citep{BFS2001}. 

The conceptual challenges discussed in this section compound with a separate set of problems: even a correctly specified model produces misleading markup estimates if the underlying data are inadequate, as Section~\ref{s:data} documents.

\section{The data constraint}\label{s:data}

The production approach needs two ingredients: a revenue share $s^j_{it}$ and an output elasticity $\gamma^j_{it}$. Data shortcomings can affect both. Mismeasured or misclassified costs---such as the COGS/SG\&A boundary discussed below---distort the revenue share directly. Missing or noisy inputs contaminate the production-function regressions used to estimate output elasticities. Estimation biases can compound direct measurement errors in shares, leaving the overall bias in estimated markups difficult to sign \textit{a priori}. This section focuses on the data problems; Section~\ref{s:econometrics} takes up some of the econometric challenges they create. Microdata have long been known to fall short of theoretical ideals \citep{Grunfeld_Griliches_1960, GrilichesRingstad}. These difficulties might help explain why markup estimates diverge so much across studies. They also point to directions for future research, as we highlight at the end of this section.

We first discuss how measurement error biases markups, then catalog specific data shortcomings, assess sample representativeness, and draw on historical parallels to offer recommendations.

\subsection{Why measurement error matters}
\label{s:measurement}
Why do data problems matter for markup estimation? The production markup $\hat{\mu}_{it}$ is the ratio of an estimated output elasticity $\hat{\gamma}^j_{it}$ to a measured revenue share $s^j_{it}$. This ratio structure means that measurement errors can distort markups in ways that are predictable in form but difficult to sign overall.

First, micro datasets consistently miss some inputs. For example, proprietary software, informal labor arrangements, and other intangibles often do not appear in production surveys. Similarly, the skill composition of the labor force or variations in factor utilization can also be considered as missing inputs. To see why missing inputs matter, suppose we know a priori that returns to scale are constant. As discussed in Section~\ref{s:growth_rates}, the markup then equals revenue divided by total cost: $\hat{\mu}_{it} = P_{it} Y_{it}/\text{Cost}_{it}$. If we miss any inputs, we understate economic costs and overstate the markup. The bias approximately equals the input's cost share times the measurement error.\footnote{Measuring input $j$ as $\hat{X}^j = (1+\epsilon)X^j$ yields markup $\hat{\mu} = \mu/(1+\epsilon c^j)$ with an error of about $-\epsilon c^j$ percent.}

Now suppose instead we estimate output elasticities econometrically, rather than imposing constant returns. Missing inputs lead to omitted variable bias. Suppose we take materials as the fully flexible input and estimate a Cobb-Douglas production function. If we omit a portion of, say, intangible capital, then all regression coefficients are potentially biased. The bias in the estimated output elasticity of materials, $\hat{\gamma}^M$, depends on the partial correlation between omitted capital and materials, conditional on included variables. Because this is a conditional correlation, the bias can be positive or negative even if the unconditional correlation is positive---so the direction is not obvious without knowing the data.

A parallel concern is whether output itself is mismeasured. We defer this question to Section~\ref{s:econometrics}, where unobserved firm-level output prices enter as the second layer of endogeneity.

Second, even when inputs are not systematically missing, they may be measured with noise. Classical measurement error in a single input attenuates its own estimated elasticity. However, the effect on the coefficients for other (accurately measured) inputs is, in general, ambiguous, depending on conditional correlations in the data. In practice, multiple inputs might be mismeasured. The net effect on output elasticities and markups depends on which inputs carry more noise and how correlated different inputs are; it cannot generally be signed without dataset-specific knowledge. For instance, suppose materials are mismeasured, leading (on its own) to attenuation bias in $\hat{\gamma}^M$. But measurement error in labor could partially or totally offset the attenuation: when noise obscures labor's true variation, ordinary least squares (OLS) assigns some of labor's productive contribution to materials, which tends to push $\hat{\gamma}^M$ back up.

One input that almost surely suffers sizeable measurement error, probably larger than the noise in labor, is capital. Book values, uncertain depreciation schedules, and missing intangibles all make capital hard to measure accurately. \citet{collard2016production} suggest that Monte Carlo simulations can help assess the quantitative (and directional) bias that arises. In their simulations, when the variance of capital's measurement error is 40\% of capital's conditional variance, capital's coefficients can be biased downward by a factor of two. But, for estimating the output elasticity for materials, the sizeable mismeasurement of capital has a more muted effect than does measurement error in labor. Capital adjusts slowly---firms cannot instantly buy or sell machines---whereas materials respond more freely to current conditions. Capital and materials therefore move together less than do labor and materials. Hence, mismeasured capital is less able to push $\hat{\gamma}^M$ upward. The upshot is that the net bias in estimated markups from capital mismeasurement is smaller than one might fear. But it is still present and its direction depends on the specifics of the data.

Nonetheless, the practical stakes are high. Even in this simplified setup, plausible measurement error levels can shift markup estimates by 5--15\% \citep{collard2016production}, though the direction depends on conditional correlations among inputs, not just on which input is mismeasured. If the variance of the measurement error is relatively constant, the bias mainly affects the level of markups. Such a bias, even if constant, could vary across inputs, as reported in Section (\ref{s:garden}). If the variance of the measurement error is changing over time, it could also influence estimated trends.

\subsection{Specific measurement shortcomings in standard datasets}
The previous subsection focused on the general mechanics of how missing inputs or classical measurement error could affect results. This subsection highlights five specific shortcomings in many datasets researchers regularly use. The shortcomings illustrate the general problems of missing inputs and classical measurement error, as well as raising additional issues. We order the five based on our subjective view of how difficult they are in practice. 

\paragraph{Revenue vs.\ quantity data are the most fundamental problem.}
Most micro datasets record revenues, not physical quantities and prices separately. In U.S.\ microdata, both Compustat and individual Census surveys offer predominantly revenue-based output measures. Separating price from quantity is possible only for a narrow slice of the economy---\citet{foster2008reallocation} study just 11 manufacturing products, including coffee, ready-mixed concrete, and motor gasoline. Such data are generally unavailable at the scale needed to study broad sectors or the whole economy. That said, survey data from some other countries---Colombia and India, for instance---sometimes separate output price from quantity. And, many customs datasets used in international trade also contain unit values.

When \citet{foster2008reallocation} do separate prices from quantities, they uncover a striking puzzle: physical productivity is \textit{inversely} correlated with price, yet revenue productivity is \textit{positively} correlated with price. The resolution reveals why revenue data creates problems for markup estimation. A physically productive firm has lower costs and can profitably charge less---hence the negative price correlation. Revenue productivity mixes efficiency with pricing power: high revenue productivity could reflect genuine efficiency or simply high markups. Without observing prices and quantities separately, we risk conflating operational excellence with market power.\footnote{Research from other countries reinforces these concerns. \citet{garcia2019exporting} show that using revenue productivity rather than physical productivity underestimates export efficiency gains in Chilean plant data. \citet{lenzu2022financial} combine firm-level output prices and quantities with quasi-experimental variation in credit supply in Belgium and show that revenue-based measures underestimate the long-run elasticity of physical productivity to credit supply by half.} We return to this challenge in Section~\ref{s:econometrics}.

\paragraph{Missing inputs and input quality create hard-to-detect biases.}
Microdata rarely contain direct information on input quality, which can bias production function estimates as well as productivity comparisons across firms \citep{Griliches1957,fox2011input, grieco2016production}. Sometimes inputs are not recorded at all. \citet{autor2020fall} study labor shares in U.S. Census microdata for six large sectors; only in manufacturing can they construct labor shares of value added, because most sectoral Census microdata lack systematic information on intermediate inputs. Even in manufacturing, inputs of business services are often missing. As discussed in the previous section, these omissions are likely to bias estimated output elasticities, whether estimated via cost-shares or regression.

\paragraph{The economics of COGS vs. SG\&A is less clear than it seems}
Compustat reports total employees and book value of capital, but lacks the detailed input classifications available in Census surveys---and intermediate inputs are missing entirely.\footnote{Compustat's staff expense variable covers only 5--20\% of firm-years depending on sample construction, with non-random selection toward larger firms and specific industries. Even \citet{brynjolfsson2003computing}, who used Compustat for production function estimation, had to impute labor costs for firms with missing staff expense data. The headcount variable is more widely available but is unaudited and misses labor quality, non-employee labor, and outsourced services.} Instead, firms classify costs under Generally Accepted Accounting Principles (GAAP) as either Cost of Goods Sold (COGS) or Selling, General, and Administrative (SG\&A) expenses. COGS covers costs directly attributed to production: direct materials, direct labor, and manufacturing overhead. SG\&A covers everything else: executive salaries, marketing, R\&D, and administrative overhead. \citet{deloecker2020rise} argue that this accounting structure offers an advantage over Census surveys: whereas Census data lump production and overhead workers together, GAAP requires firms to separate production costs from overhead, providing a principled split that standard input classifications cannot. They treat COGS as the sole flexible input and exclude SG\&A entirely from the production function. In their data sample, the COGS share of revenue has trended down in recent decades, while their estimated output elasticity for COGS does not change much. They interpret the declining revenue share as implying rising markups.

But the mapping between accounting categories and economic inputs is not clear-cut. Even if SG\&A is partly quasi-fixed rather than fully variable, omitting a productive input from the production function biases the estimated elasticity of the remaining inputs upward---just as omitting capital does. The bias worsens as the omitted input's share of costs rises, and the SG\&A share of operating expenses has grown steadily. \citet{traina2018aggregate} shows this matters enormously: including SG\&A with COGS as the flexible input eliminates most of the rise in estimated markups since 1980. Within-firm changes in SG\&A are approximately smooth across ranks \citep{traina2018aggregate}, suggestive of routine expenses that vary with output rather than the lumpy adjustments expected of fixed costs.

The stakes are high. As documented in Section~\ref{s:garden}, a firm with \$70 million in COGS and \$20 million in SG\&A generating \$100 million in revenue faces a flexible input share of either 0.70 (COGS only) or 0.90 (COGS+SG\&A). Assuming an output elasticity $\gamma = 1$, the implied markups are either 43\% or 11\%. Whether the declining COGS share reflects changing production economics or evolving accounting practices remains an open question whose resolution, as \citet{JorgensonGriliches1967} emphasized, requires understanding the data before interpreting the results.

\paragraph{Capital measurement is a well-known but only partially addressable problem.}
Micro datasets typically provide book values of capital---historical purchase prices adjusted by accounting depreciation rules---rather than economic values reflecting productive capacity. Book values can diverge dramatically from true productive input: a fully depreciated but still-productive machine carries zero book value; a recently purchased but obsolete asset maintains a high one \citep{becker2006micro}.

The problem deepens with intangible assets. A growing literature documents the importance of intangible capital---brand equity, organizational capital, and other assets beyond software and R\&D \citep{corrado2009intangible}---but these typically appear in accounting data as operating expenses rather than capitalized assets.\footnote{\citet{KohSantaeulaliaZheng2020} argue that the measured decline in the U.S.\ labor share is entirely accounted for by the capitalization of intellectual property products in the national accounts---illustrating how changes in accounting treatment, rather than changes in market power, can shift measured factor shares.} \citet{karabarbounis2019accounting} document substantial growth in ``factorless income''---the apparent profits remaining after subtracting measured payments to labor and capital from GDP. This could reflect true market-power rents, an understated user cost of capital, or returns to unmeasured intangible capital.

Under the cost-share approach, any understatement of true costs translates directly into inflated markup estimates. Econometric corrections can help: \citet{lizalgaluscak2012}, \citet{collard2016production}, and \citet{kimpetrin2016} show that correcting for capital mismeasurement can nearly double estimated capital elasticities. The cost-share approach also requires a measure of the user cost of capital: because most capital is owned rather than rented, the user cost must be imputed from depreciation rates, expected capital gains, risk premia, and tax adjustments \citep{hall1967tax, JorgensonGollopFraumeni1987}. Getting this right is itself a major measurement challenge.

\paragraph{Measuring the cost of inflexible inputs requires prices that are rarely observed.} For inputs that cannot be freely adjusted---capital and, in the short run, certain types of labor---the relevant cost concept is the shadow price, not the market price. When a firm cannot instantly hire or fire workers, or cannot costlessly buy or sell capital, the shadow value of those inputs reflects adjustment costs and future expectations, not just today's market wage or rental rate. Using observed market prices to measure the cost of inflexible inputs therefore introduces systematic error into both the revenue share and the estimated elasticity simultaneously, that is, into the two ingredients whose ratio defines the markup. This problem cuts across all of the measurement challenges above, since capital and quasi-fixed labor are precisely the inputs most likely to be mismeasured in other dimensions as well \citep{fox2011input, ornaghi2006, KimTraina2025}.

\subsection{Representativeness: are we capturing the right firms?}

A key selling point of the production approach is that it can be applied at scale. In this regard, even perfect measurement at the firm level would not resolve the challenge that the firms we observe most comprehensively might not be representative of the economy we want to understand.

The U.S. Census of Manufactures is the gold standard for production function estimation: broad coverage, detailed input data, and separation of plants from firms \citep{dunne1988patterns, baily1992productivity, foster2008reallocation, syverson2011determines}. But manufacturing now accounts for only about 12\% of U.S. private-industry value added. If markups in services or finance differ systematically from those in manufacturing, studies based on Census manufacturing data are learning about a shrinking and un(der)representative slice of the economy. The problem is not just coverage: manufacturing firms have relatively clear distinctions between production workers and overhead, and between raw materials and administrative expenses. Service firms blur these boundaries. A software engineer at a technology firm: production labor or overhead? As the economy shifts toward services, accounting categories designed for manufacturing become less economically meaningful. Census microdata remain best-in-class in coverage and sample size, and researchers are exploiting within-sector detail---for example, \citet{DhyneEtAl2022} use Belgian firm-to-firm transaction data that capture input linkages and allow direct observation of intermediate input prices---but the representativeness problem does not go away.

Given manufacturing's declining share, researchers increasingly turn to Compustat, which offers broad sectoral coverage and is easy to access. Compustat covers only public firms, which account for roughly half of industry sales \citep{davis2007volatility}, and some 40 percent of those sales come from foreign operations, complicating the interpretation of domestic market power. Several papers \citep{ali2008limitations, Keil2017trouble, DeckerWilliams2023concentration} find that common concentration measures at the 4-digit NAICS level based on Compustat are only weakly correlated with those from the more comprehensive economic census. Sample construction choices also matter: \citet{BenkardMillerYurukoglu2025} show that unreported sample restrictions in \citet{deloecker2020rise} drop 27\% of available Compustat observations. With the full sample, they find that markup increases are more muted and concentrated in Finance and Insurance, where cost measurement is especially problematic.

Whatever the dataset, a basic diagnostic is to check whether micro aggregates match published national accounts totals. \citet{FernaldPiga2023} find that factor shares in the U.S.\ Census of Manufactures are wildly different from those in the national accounts: in 1977, labor's share of value added in manufacturing plants was only 31\%, compared to 58\% in the BEA-BLS KLEMS dataset for the same sector. Such large discrepancies in factor shares translate directly into different markup estimates---what looks like a high markup might just be missing labor compensation or business services. The gap between micro data and national accounts reflects more than coverage differences: national accounts integrate multiple data sources---tax records, banking surveys, innovation surveys---that researchers rarely access, and micro data should not be expected to aggregate to published totals without understanding that integration.

\subsection{Historical lessons and recommendations}
Measurement challenges are not new and are, of course, pervasive. How economists confronted earlier episodes offers lessons for the production approach to markups.

Both macroeconomics and industrial organization were confronted with residuals that potentially conflated true economic phenomena with measurement artifacts. In macroeconomics, the puzzle was aggregate total factor productivity. \citet{solow1957technical} found that (what we now call) the Solow residual, explained 88 percent of growth in U.S.\ output-per-hour growth from 1909 to 1949. \citet{JorgensonGriliches1967} argued that much or most of this residual 88 percent largely reflected measurement error in output and inputs. They issued what has been called a ``manifesto'' for improving measurement \citep[][p.~579]{BermanJaffe_2024}.\footnote{Measurement progress since the 1960s owes much to Jorgenson and Griliches, whose contributions extend well beyond their joint work. Griliches emphasized microdata, while Jorgenson focused on macrodata and industry aggregates \citep{BermanJaffe_2024, fernald_2024}. Despite these advances, their hypothesis that total factor productivity is merely measurement error falls short: from 1948--2024, total factor productivity still accounts for roughly half of labor-productivity growth \citep{fernald_quarterly}.} They applied index-number theory to construct better measures: Divisia output indexes, hedonic price adjustments, theory-based indexes of labor composition, and user-cost weights for heterogeneous capital. More than half a century of research in national accounting circles has refined these measures at an industry and aggregate level. Such data are not just the sum over firm-level datasets, because they combine multiple data sources and allow statistically representative surveys. This is a reason why the growth-rate implementation of production-approach markups has mostly used industry data (Section~\ref{s:growth_rates}).

Industrial organization faced a parallel crisis. The ``structure-conduct-performance'' (SCP) paradigm \citep{bain1951relation, bresnahan1989empirical, schmalensee1989inter} tried to infer market power from correlations between industry concentration and accounting profit rates. It failed on two fronts: causal ambiguity \citep{demsetz1973industry} (even with perfect data, the correlation between concentration and profits is uninterpretable without additional economic structure) and the gap between accounting profitability (at best a markup of price over average variable cost) and true economic marginal cost \citep{fisher1983misuse}. The same capital measurement problems that bedevil the production approach---risk premia, inflation, depreciation rules, missing intangibles---made accounting profitability an unreliable proxy for true economic returns. The IO response was to abandon broad cross-industry regressions in favor of deep, structurally disciplined studies of specific industries, using institutional knowledge to pin down the assumptions that aggregate approaches cannot test \citep{berry2019increasing}.

The production approach to markups tries to be broad like the macro tradition and agnostic about the structure of demand like early IO---and in doing so it has to face the measurement challenges of both. Like the SCP paradigm, it is sometimes forced to use accounting costs and book values. Like early aggregate productivity accounting, it must take a stand on input measurement without the cross-checking that comes from integrating multiple data sources. The implication is not that the approach is hopeless, but that substantive progress can require both better data---the Jorgenson-Griliches response---and more disciplined structural assumptions---the IO response.

Four practical recommendations follow from this history. First, the Jorgenson-Griliches agenda of improving underlying data remains as relevant today as in 1967. Better measurement of inputs---especially capital services, intangible investments, firm-level prices, and the boundary between production and overhead labor---would strengthen the foundation for production-based markup estimation \citep{griliches1998production, lev2001intangibles}. Second, because the direction of bias from classical measurement error is ambiguous---as discussed in Section~\ref{s:measurement}---simulation can help assess the seriousness of a problem in a given application. \citet{collard2016production} is an example: They conduct dataset-specific simulations to bound measurement biases before drawing conclusions about markup levels or trends. Third, an easy check is to report how micro aggregates compare to published national accounts totals---large discrepancies are a warning sign that something important is being missed, as \citet{FernaldPiga2023} illustrate for manufacturing. Fourth, when measurement problems cannot be resolved, researchers should be explicit about the resulting bias and whether it can plausibly explain the patterns attributed to market power. These measurement problems also feed directly into the econometric challenges of estimating output elasticities, which Section~\ref{s:econometrics} takes up next.

\section{Practice and pitfalls in estimating output elasticities}
\label{s:econometrics}
Production-based markups are only as reliable as the output elasticities that underpin them. The measurement challenges documented in Section~\ref{s:data}---particularly unobserved output prices---feed directly into this estimation problem. The challenge is that most tools for estimating those elasticities were built for a world of perfect competition, and using them to measure departures from perfect competition creates problems that run deeper than is commonly appreciated. The applied economist has a rich toolbox, but its tools force tradeoffs: some are restrictive for an IO economist who wants to remain agnostic about the structure of production and competition; others are overly general for a macroeconomist who wants to calibrate a specific model.

We organize the discussion in four parts. First, we build up the estimation problem in two steps, showing how moving from physical output to revenue data adds unobserved prices to the regression residual and creates a new source of bias. Second, we show what goes wrong if those extra terms are ignored---including a striking result that naive estimation with revenue data can be entirely uninformative about markups. Third, we evaluate the main econometric tools the literature has developed, asking how each fares once we take the richer residual seriously. Fourth, we offer an assessment of what the toolkit can and cannot deliver.

\subsection{The problem: two layers of endogeneity}
The fundamental challenge in estimating output elasticities is endogeneity: firms choose their inputs in response to forces the econometrician cannot observe, so input choices are correlated with the regression residual. This problem has two layers in realistic settings, and understanding each layer separately is the key to understanding why estimating output elasticities is so challenging.

\paragraph{Layer 1: Physical productivity.}
The cleanest statement of the problem starts from the physical production function. For concreteness, take the Cobb-Douglas production function \eqref{e:pf} with three inputs and a firm-specific productivity term $a_{it}$:
\begin{equation}\label{e:pf_layer1}
    y_{it} = \gamma^K k_{it} + \gamma^L l_{it} + \gamma^M m_{it} + a_{it},
\end{equation}
where all variables are in logs and $\gamma^X$ denotes the output elasticity of input $X$. The residual here is just productivity $a_{it}$. The endogeneity problem is the classic ``transmission bias'' \citep{marschak1944random}: a firm that draws a high productivity shock may want to use more materials, so $m_{it}$ and $a_{it}$ move together. OLS attributes that comovement to technology and overstates $\hat{\gamma}^M$.\footnote{Henceforth, we assume the bias is positive. It need not be. Suppose firms have market power, prices are sticky, and all firms receive the same productivity improvement. Marginal cost falls but firms do not adjust prices. The rising markup dampens input demand, leaving the overall OLS bias ambiguous.} Because the production approach calculates markups as $\hat{\mu}_{it} = \hat{\gamma}^M / s_{it}^M$, overstating $\hat{\gamma}^M$ directly overstates measured markups. This first layer is well understood, and the econometric toolkit was largely built to address it.

\paragraph{Layer 2: Unobserved output prices.}
Few datasets contain firm-level output quantities. Instead, researchers observe firm revenue $r_{it} = p_{it} + y_{it}$, where $p_{it}$ is the firm's (log) output price. Adding $p_{it}$ to both sides of \eqref{e:pf_layer1} gives the revenue production function that is actually estimated:
\begin{equation}\label{e:rpf_layer2}
    r_{it} = \gamma^K k_{it} + \gamma^L l_{it} + \gamma^M m_{it} + \underbrace{a_{it} + p_{it}}_{\text{residual}}.
\end{equation}
The residual now contains the firm's output price as well as its productivity. This matters because output prices are correlated with input choices for a different reason than productivity: firms with more market power charge higher prices, restrict output, and therefore use fewer inputs. Under imperfect competition, \citet{klette1996inconsistency} show that output prices and input quantities are negatively correlated, biasing output elasticities \textit{downward}---the opposite direction from Layer~1. The two biases need not cancel. More fundamentally, the revenue-based residual $a_{it} + p_{it}$ conflates technology and demand-side forces in a way that standard tools, designed for \eqref{e:pf_layer1}, were not built to handle. \citet{deloecker2016prices} confirm this is not a theoretical curiosity: using Indian manufacturing data where firm-level prices are observed, they find that revenue-based and quantity-based markup estimates diverge substantially. 

An additional complication arises when input quantities are also unobserved. Most datasets record only input expenditures $e_{it}^M = w_{it}^M + m_{it}$, so firm-specific input prices $w_{it}^M$ enter the residual as well.\footnote{The composite residual becomes $\varepsilon_{it} = a_{it} + p_{it} - \gamma^M w_{it}^M$. If firms with market power in input markets face lower prices and purchase more, input expenditures are correlated with the input-price term---a source of endogeneity discussed further in Section~\ref{s:right_FOC}.} We focus primarily on the first two layers---productivity and output prices---which are likely the dominant sources of bias in practice. 

\paragraph{A unifying look at input demand.}
To see why all three components cause endogeneity, it helps to look at what determines a firm's materials choice. Under Cobb-Douglas production, the first-order condition for materials gives their conditional demand:
\begin{equation}\label{e:materials_demand}
    m_{it} = \frac{1}{1-\gamma^M}\left(a_{it} + p_{it} + \log\gamma^M 
    - w_{it}^M - \log\mu_{it} + \gamma^K k_{it} + \gamma^L l_{it}\right).
\end{equation}
Every component of the composite residual---productivity $a_{it}$, output price $p_{it}$, and input price $w_{it}^M$---drives materials demand. That is the root of the problem: the very flexibility that makes materials a good candidate for the production-approach FOC ensures that materials respond to forces the econometrician commonly cannot observe.

\subsection{What goes wrong if you ignore the extra layers}
Ignoring the richer residual has serious consequences, even for sophisticated estimation strategies. This section traces those consequences before turning to potential fixes.

Even setting aside output and input prices, OLS applied to \eqref{e:pf_layer1} overstates $\hat{\gamma}^M$ because productive firms use more materials. Since markups are computed as $\hat{\mu}_{it} = \hat{\gamma}^M / s_{it}^M$, the overstatement flows directly into estimated markups. This is the transmission bias, and it is the reason the literature moved away from OLS toward more sophisticated approaches. The deeper problem is that moving to revenue data in \eqref{e:rpf_layer2} adds a second source of bias that can partially offset or further amplify the first, in ways that are difficult to sign \textit{ex ante}.

The most striking consequence of ignoring Layer~2 is a result due to \citet{bond2021some}:\footnote{\citet{desouza2009estimating} derives the same result in an earlier and less widely known paper.} if output elasticities are estimated from revenue data without correcting for omitted prices, the production-approach markup formula is tautologically uninformative. The logic is direct. With revenue data, OLS recovers not the output elasticity $\gamma^M$ but a revenue elasticity $\hat{\gamma}^{\text{revenue},M} = \gamma^M/\mu_{it}$---a ratio that already embeds the markup. Plugging this into the markup formula then gives:
\begin{equation*}
    \hat{\mu}^{\text{revenue}}_{it} = \frac{\hat{\gamma}^{\text{revenue},M}}{s^M_{it}} 
    = \frac{\gamma^M/\mu_{it}}{\gamma^M/\mu_{it}} = 1.
\end{equation*}
The procedure mechanically returns markups of one, regardless of its true value. 

One response to missing firm-level prices is to deflate revenues by an industry price index $p_t$. \citet{hashemi2022production} show that this ``fix'' resolves the problem only under perfect competition---where markups equal one and the exercise is unnecessary. To see the logic, the price deviation in the residual becomes $\tilde{p}_{it} = p_{it} - p_t$. 
 Under perfect competition, the deflation works because firms charge the same quality-adjusted price, so $\tilde{p}_{it}$ is zero and, thus, uncorrelated with inputs. But under imperfect competition---the case of interest---firm-level prices reflect market power and are correlated with input choices. \citet{klette1996inconsistency} show this biases elasticity estimates downward, and \citet{deloecker2016prices} show the bias is empirically meaningful. Revenue productivity may still be a useful object in some settings \citep{foster2008reallocation}, but it is not physical productivity.

Both layers of the residual create a fundamental circularity: estimating output elasticities requires controlling for markups (because markups affect input demand through \eqref{e:materials_demand}); but markups are what we are trying to measure using those elasticities. This means that any estimation strategy that ignores markups is potentially misspecified in a way that contaminates the object of interest. The circularity also explains why firm-level price data are so valuable: they allow researchers to separate the productivity and price components of the residual directly, breaking the loop. Such data are increasingly available in manufacturing and trade datasets for homogeneous goods, though services and differentiated products remain challenging.

\subsection{The toolkit: what each approach handles and what it does not}

The literature has developed a range of tools to address endogeneity in production function estimation. Each was designed primarily for Layer~1---a world where the residual is just productivity. We evaluate each tool by asking: how well does it travel to the richer-residual world of \eqref{e:rpf_layer2}?

\paragraph{Fixed effects}remove time-invariant components of the residual, but leave the time-varying components that matter most for markup estimation. Introduced by \citet{mundlak1961empirical} for agricultural production, where productivity largely reflects time-invariant characteristics like soil fertility, firm fixed effects absorb the permanent part of productivity, output prices, and input prices---a genuine contribution. But time-varying productivity shocks, demand fluctuations, and input-price variation all remain in the residual and remain correlated with input choices.

Fixed effects also exacerbate a separate problem: measurement error in capital. Because within-firm variation in capital is dominated by investment and depreciation, both of which are measured with error, demeaning amplifies that noise. \citet{griliches1998production} document that panel methods applied to micro data have produced disappointing results, with unreasonably low capital coefficients and implausible returns to scale. \citet{becker2006micro} show in an in-depth study of capital measurement that different ways of measuring capital which ought to be equivalent---such as perpetual inventory methods or inferring capital from capital-producing sectors---lead to meaningfully different production function estimates. Measurement concerns were the primary reason fixed effects were largely abandoned for production function estimation \citep{deloecker2012markups}.

\paragraph{Instrumental variables}are the most conceptually transparent solution, but the bar for a valid instrument is very high in the richer-residual world. A valid instrument $z_{it}$ must be correlated with the input choice but uncorrelated with \textit{all} components of the composite residual: productivity and output prices at minimum, and input prices when expenditures proxy for quantities. The input demand equation \eqref{e:materials_demand} clarifies what could work in principle: any variable that shifts input demand without entering the production function. Input prices $w_{it}^M$ are the natural candidate---when available at the firm level, they can isolate cost-driven variation in inputs \citep{doraszelski2013r}. But firm-level input prices are often weak or correlated with local demand conditions that also affect output prices, undermining validity \citep{gandhi2020identification}.\footnote{Even if a firm takes input prices as given, general equilibrium effects can create problems. If all firms in a market receive a positive productivity shock, the market price for their output may fall, creating a correlation between the shock $a_{it}$ and the output price $p_{it}$ that contaminates the residual.} Aggregate demand instruments---government spending, oil price shocks, monetary policy---have been used in industry time-series approaches (Section~\ref{s:growth_rates}). But they face similar concerns if demand shocks simultaneously shift costs or induce entry and exit. Despite these limitations, instrument relevance is testable, and IV deserves serious consideration when strong instruments are available: its assumptions are transparent in a way that alternatives are not.

\paragraph{Control functions}are the current workhorse of the production function literature, largely because they avoid the need for external instruments. However, their key assumptions are individually questionable under imperfect competition and must hold simultaneously. The approach, developed by \citet{olley1996dynamics} and extended by \citet{levinsohn2003estimating}, uses observable firm decisions as proxy variables for unobserved productivity. \citet{olley1996dynamics} originally used investment as the proxy. But investment is often lumpy and infrequent. Hence, \citet{levinsohn2003estimating} proposed intermediate inputs, which adjust more smoothly and respond more directly to productivity shocks. The idea is that if a firm's investment or materials choice responds monotonically to productivity, we can invert that relationship to express unobserved productivity as a function of observables:
\begin{equation}\label{e:materials_control}
    a_{it} = h^{-1}(k_{it}, l_{it}; m_{it}).
\end{equation}
Substituting this into the production function removes the endogenous productivity term. In practice, the approach proceeds in three steps:
\begin{enumerate}
    \item \textit{First stage}: Regress output on inputs and the control variable (typically with a flexible polynomial) to estimate the combined production and control function $\phi_t(k_{it}, l_{it}, m_{it})$. The residual from this regression is interpreted as measurement error. The fitted value $\hat{y}_{it}$ provides output cleansed of classical measurement error. Individual production function parameters are not yet identified, since the materials coefficient cannot be separated from the part of $h^{-1}(\cdot)$ that depends on materials.
    \item \textit{Second stage}: For candidate parameters $(\gamma^K, \gamma^L, \gamma^M)$, back out productivity as $\hat{a}_{it}(\gamma) = \hat{y}_{it} - \gamma^K k_{it} - \gamma^L l_{it} - \gamma^M m_{it}$.
    \item \textit{Third stage}: Productivity is assumed to be Markovian. Hence, we can regress $\hat{a}_{it}$ on $\hat{a}_{i,t-1}$ to decompose it into a predictable component and an innovation $\xi_{it}$. Because $\xi_{it}$ is orthogonal to inputs chosen before the innovation is realized, lagged inputs form the basis of moment conditions for generalized method of moments (GMM) estimation: $\mathbb{E}[\xi_{it}(\gamma) \cdot z_{i,t-1}] = 0$. With revenue data, however, this step recovers revenue productivity $a_{it} + p_{it}$ rather than physical productivity, and the Markov assumption must hold for the composite---a strong requirement when demand and technology respond to different forces.
\end{enumerate}
\citet{ackerberg2015identification} refine the approach by assuming labor is chosen before materials, addressing collinearity between the proxy and the flexible input. But identification remains fragile. The output elasticity of a flexible input---the object needed for markup inference---is underidentified in many standard settings because the proxy's dual role (absorbing productivity \textit{and} generating identifying variation) creates collinearity: when materials optimally adjust to productivity, they become collinear with it, leaving no independent variation to identify how inputs transform into output. \citet{gandhi2020identification} show these identification concerns are even more severe in gross-output settings, where intermediates enter both the production function and the control. \citet{flynn2019measuring} extend the analysis to settings with markups, showing that lagged flexible instruments can generate spurious skewness in the estimated markup distribution; imposing constant RTS resolves this identification failure.

Three assumptions are load-bearing---and each is strained by imperfect competition. First, what is known as the \textit{scalar unobservable} assumption requires productivity to be the \textit{only} unobserved driver of the proxy variable. But in the richer-residual world, input choices also respond to output prices and markups, so the proxy can reflect multiple unobservables and not be invertible. For instance, \citet{ackerberg2007econometric} show that when there are two latent state variables---productivity and an unobserved markup determinant---two independent controls are needed, not one. Second, \textit{monotonicity} requires the proxy to be strictly increasing in productivity, but when more productive firms charge lower prices (the Klette-Griliches logic), revenue productivity $a_{it} + p_{it}$ may not be monotonic in $a_{it}$. The monotonicity assumption can also fail for macroeconomic reasons: with sticky prices or inelastic short-run demand, firms may initially \textit{reduce} flexible input use following a positive productivity shock and only expand later, as documented by \citet{Gali1999} and \citet{BFK2006technology}. \citet{doraszelski-Li-invertibility} show that invertibility is testable and propose modifications when it fails. Third, the \textit{Markov} assumption on productivity must now hold for the composite residual $a_{it} + p_{it}$, which requires productivity and prices to evolve in lockstep---implausible when demand responds to different forces than technology.

Proposed remedies tend to reintroduce the structural assumptions the production approach was designed to avoid. \citet{doraszelski2019using} show imperfect competition renders the control function uninvertible when input choices depend on unobserved markup determinants. One response is to add market shares as a second control variable \citep{deloecker2020rise,GrassiMorzentiRidder2022Hitchhiker}, but this requires assumptions about the competitive environment. \citet{kirov2026measuring} advocate including observable price controls directly in the first-order conditions \citep[see also][]{kasahara2020nonparametric,AckerbergDeLoecker2024}. These strategies capture price variation that drives input choices, but their validity rests on assumptions about firm behavior---precisely the structural detail the production approach aims to avoid. These problems concern whether control functions succeed on their own terms---but there is a more basic gap. Even when the proxy assumptions hold, the method addresses Layer~1 (unobserved productivity) without resolving Layer~2: inverting materials demand recovers $a_{it}$, not $p_{it}$, so the Bond et al.\ uninformativeness problem remains unless firm-level prices or an explicit demand model supplement the control function. Control functions remain the workhorse; researchers should be explicit about which layers of the residual they address and which they leave unresolved. At minimum, researchers should report whether their proxy addresses output-price variation, input-price variation, or both, and conduct sensitivity checks---such as the invertibility test of \citet{doraszelski-Li-invertibility}---to assess whether the scalar unobservable assumption is plausible in their setting.

\paragraph{Dynamic panel methods} replace the scalar unobservable assumption with time-series assumptions, but those assumptions must now hold for the full composite residual. These methods difference the production function to remove fixed effects and use lagged inputs as instruments for current input changes, relying on the assumption that lagged inputs are uncorrelated with productivity innovations \citep{arellano1991some, blundell1998initial, blundell2000gmm}. System GMM combines differenced and level equations to improve efficiency. With revenue data, however, the error term becomes $\Delta a_{it} + \Delta p_{it}$, so the moment conditions require both productivity and prices to follow similar time-series processes---a demanding and implausible
requirement when demand and technology respond to different forces. \citet{bond2005adjustment} also note that lagged instruments only have identifying power in the presence of frictions such as adjustment costs, but adjustment costs violate the static first-order condition that underpins the markup calculation itself. \citet{brand2019estimating} proposes treating observed output as a noisy signal of productivity, allowing unobservables to evolve nonlinearly, and using lagged output as an instrument when productivity is persistent and measurement error is not. A valuable validation would use datasets that separate output prices from quantities to test whether physical and revenue productivity have time-series properties consistent with the approach---those assumptions underpin the estimated elasticities and hence any markups inferred from them. These refinements are promising, but the fundamental tension between the time-series assumptions and the composite residual remains.

\paragraph{Structural demand modeling} is the most direct solution to the omitted-price problem, but it sacrifices the production approach's model-free appeal. If we are willing to specify a demand system, we can derive estimating equations in terms of observed revenue rather than unobserved quantities. For example, \citet{klette1996inconsistency} propose combining the production function with an isoelastic CES demand curve:
\begin{equation}\label{e:ces_demand}
    y_{it} = y_t - \eta(p_{it} - p_t),
\end{equation}
where $\eta$ is the (absolute value of the) price elasticity of demand and $y_t$, $p_t$ are industry (not firm) quantity and price indexes. Substituting into the production function yields a revenue-based estimating equation whose coefficients are combinations of output elasticities and the demand elasticity:
\begin{equation}\label{e:revenue_structural}
    r_{it} - p_t = \frac{\eta-1}{\eta}\left(\gamma^K k_{it} + \gamma^L l_{it} + \gamma^M m_{it}\right) + \frac{1}{\eta} y_t + \frac{\eta-1}{\eta} a_{it}.
\end{equation}
Output elasticities can then be separated from demand elasticities and estimated using control function methods. \citet{deloecker2011product} extend this approach to multiproduct firms; \citet{ruzic2021returns} and \citet{CLRS:2024} extend it to models with heterogeneous markups and oligopolistic competition. The benefit is transparency: structural assumptions are explicit rather than buried in econometric corrections, and a fuller model supports richer counterfactual analysis. The cost is that markups become functions of the assumed demand and conduct model---returning, in effect, to traditional IO methods. If demand and competition must be specified to estimate output elasticities, the production approach's central appeal---avoiding those specifications---is substantially weakened.

These methods face the same fundamental tension: the FOC that delivers $\mu = \gamma/s$ assumes a flexibly adjusted input, but identifying $\gamma$ often requires frictions or structure that violate this very assumption.

\subsection{Recommendations for applied work}
No existing method fully resolves the endogeneity problem. That is worth stating plainly. The production approach's promise of estimating markups without specifying market structure remains, in its most ambitious form, unfulfilled. What follows is practical guidance given that reality.

First, firm-level price data are the cleanest fix when available. Collecting firm-specific output and input prices can eliminate the output-price bias directly, rather than routing around it econometrically. Manufacturing and trade datasets increasingly contain quantity information for homogeneous goods, allowing researchers to construct firm-level prices. \citet{GrassiMorzentiRidder2022Hitchhiker} provide the most systematic assessment using French manufacturing data, comparing revenue-based and quantity-based markup estimates. Their simulations---calibrated to a specific oligopoly model (both Cournot and Bertrand cases)---yield a 0.93 correlation between revenue-based and true log markups. But in the French data, the correlation between revenue-based and quantity-based markups is only 0.3, well below the simulation benchmark. If that gap is representative, conclusions about markup trends and dispersion drawn from revenue data alone should be treated with caution. Services and differentiated products remain challenging, but where quantity data exist, they should be used. For differentiated products, combining scanner data with production records---as in \citet{GrassiMorzentiRidder2022Hitchhiker}---can approximate firm-level prices even when true quantities are unavailable. Expanding such matched datasets should be a priority for the field.

Second, instrumental variables deserve serious consideration when strong instruments are available. IV is the most assumption-transparent method in the toolkit: the exclusion restriction is explicit and testable, unlike the scalar unobservable and monotonicity assumptions underlying control functions. When firm-level input prices are available and strong, they provide a natural instrument that addresses both layers of endogeneity simultaneously. The main caveat is factor market power: if firms with monopsony power in labor or materials markets face systematically lower input prices, input prices are correlated with the input-price term in the residual, violating the exclusion restriction. 

Third, control functions remain the workhorse, but researchers should be explicit about what they leave unresolved. The approach is valuable precisely because it requires fewer external instruments than IV, and its practical implementation is well-developed \citep{olley1996dynamics, levinsohn2003estimating, ackerberg2015identification}. But in settings with meaningful price dispersion or market power, the scalar unobservable assumption is likely violated, the monotonicity assumption is strained, and the Markov assumption on the composite residual is implausible. Researchers using control functions should report sensitivity to these assumptions and, where possible, test invertibility following \citet{doraszelski-Li-invertibility}.

Fourth, the applied economist should be honest about the gap between the toolkit and the problem. Each method makes different compromises between identification strength and the assumptions underlying markup inference. Fixed effects and deflation work only under restrictive conditions. IV needs variation uncorrelated with all three residual components---a high bar. Control functions assume productivity is the sole unobservable, failing when demand affects input choices. Dynamic panels rely on implausible joint time-series assumptions for the composite residual. Structural demand modeling addresses these issues by imposing explicit structure, but sacrifices the model-free appeal of the production approach. The right choice depends on the research question, the institutional setting, and the data available. What should not vary is the obligation to be transparent: state clearly which sources of bias the chosen method addresses, and which it does not. Acknowledging these limits honestly is the foundation; Section~\ref{s:Call_to_Arms} builds on it with concrete recommendations for making production-based markup estimation more reliable and more transparent.

\section{A Call to Arms}\label{s:Call_to_Arms}
We began this review with a unifying FOC linking revenue shares, output elasticities, and markups. We end by using that same FOC to launch three methodological calls to arms. First, a transparency decomposition quantifying how much revenue-share variation reflects markups versus output elasticities. Second, systematic comparisons of production- and demand-based markups. Third, more work mapping firm-level markup heterogeneity into macroeconomic models.

\subsection{Transparency: An R-squared for revenue shares}
A recurring question in this review is: how much of the observed variation in revenue shares---across firms and over time---reflects demand-side forces like markups, and how much reflects technological differences like output elasticities?

Implementations of the production approach frequently impose strong parametric restrictions on output elasticities. As a result, variation in revenue shares gets disproportionately loaded onto markups. To see why, take logs of FOC \eqref{e:FOC-ratio} for input $X^j$:
    \begin{align}\label{e:r-squared-regression}
     \ln s^j_{it} &= \ln\gamma_{it}^j-\ln\mu_{it}.
     \end{align}
This identity says that every firm-level deviation in a revenue share must be accounted for by either the output elasticity or the markup. The variance decomposition makes this explicit:
    \begin{align*}
        \text{Var}(\ln s^j_{it}) = \text{Var}(\ln\gamma_{it}^j) + \text{Var}(\ln\mu_{it}) - 2\,\text{Cov}(\ln\gamma_{it}^j, \ln\mu_{it}).
    \end{align*}
When estimation restricts how much output elasticities can vary---compressing $\text{Var}(\ln\widehat\gamma_{it}^j)$---markups can inherit the leftover variation by construction. The concern then is that parametric restrictions on output elasticities mechanically inflate the dispersion of estimated markups and hence markups' explanatory power for revenue shares.

Consider this stark Cobb-Douglas example. Suppose we estimate a single output elasticity $\widehat\gamma^j$ shared by all firms in a given period. Then $\text{Var}(\ln\widehat\gamma_{it}^j)=0$ and $\text{Cov}(\ln\widehat\gamma_{it}^j, \ln\widehat\mu_{it})=0$ by construction, and the identity collapses to:
    \begin{align*}
        \text{Var}(\ln s^j_{it}) = \text{Var}(\ln\widehat\mu_{it}).
    \end{align*}
All revenue-share variation is attributed to the production-based markup. A regression of $\ln s^j_{it}$ on $\ln\widehat\mu_{it}$ would yield an $R^2$ of exactly one---not necessarily because true markups explain all the variation, but because the parametric restriction on output elasticities forced all the variation of revenue shares into the residual claimant, the production-based markup. An $R^2$ of one means the decomposition has no testable implications: it cannot distinguish a world where markups truly explain all variation from one where a restrictive functional form simply leaves no room for output elasticities to explain any.

\textbf{A simple $R^2$ diagnostic.} Rather than requiring a full variance decomposition, a simple regression offers a practical transparency check. Consider:
	\begin{align*}
	\ln s^j_{it} = \alpha + \beta\ln\hat{\gamma}^j_{it}+\varepsilon_{it},
	\end{align*}
with $\alpha$ the constant, $\beta$ the slope coefficient and $\varepsilon_{it}$ the error term. The $R^2$ of this regression measures the share of revenue-share variation linearly explained by the estimated output elasticity. The remaining $1 - R^2$---the residual---captures the share attributed to the estimated markup. Consistent with the rest of this review, the markup is the residual: whatever revenue-share variation the output elasticity cannot explain. Researchers can also run this within dimensions of interest (year, industry, or industry-year) by including fixed effects; the within $R^2$ then quantifies relative explanatory power conditional on those controls.

Table \ref{table:R2} applies this diagnostic to Compustat data using replication code from existing papers. The Output Elasticity columns report the $R^2$ from the above regression; the Markup columns report $1-R^2$. All estimates correspond to the headline specification in \citet{deloecker2020rise}, which uses Cost of Goods Sold (COGS) as the flexible input.

\begin{table}[H]
\renewcommand{\thetable}{2}
\centering
\caption{An $R^2$ Decomposition of Revenue-Share Variation: Output Elasticities vs Markups}\label{table:R2}
\scriptsize\singlespacing
\renewcommand{\arraystretch}{0.84}
\begin{tabularx}{\linewidth}{p{2.2in} C C C C}
\multicolumn{1}{l}{} & \multicolumn{2}{c}{\textbf{Full Sample}} & \multicolumn{2}{c}{\textbf{Industry-Year}} \\
\cmidrule(r){2-3} \cmidrule(r){4-5}
& Output Elasticity & Markup & Output Elasticity & Markup \\
\midrule
\multicolumn{2}{l}{\textbf{De Loecker, Eeckhout \& Unger (2020)}} & & & \\
\textit{Replication File} & & & & \\
\addlinespace[0.5em]
Cobb-Douglas, Industry-Year & 0.04 & 0.96 & 0.00 & 1.00 \\
\addlinespace[2pt]
\textit{Cost Shares under Constant RTS} & & & & \\
\addlinespace[0.5em]
Cost Share, 2-digit NAICS & 0.00 & 1.00 & 0.00 & 1.00 \\
Cost Share, 4-digit NAICS & 0.01 & 0.99 & 0.00 & 1.00 \\
\addlinespace[2pt]
\multicolumn{2}{l}{\textbf{De Ridder, Grassi \& Morzenti (2024)}}\\
\multicolumn{2}{l}{\textit{Markup Toolbox: Translog, 2nd order}} & & \\
\addlinespace[0.5em]
No interactions of inputs, Firm-Year  & 0.01 & 0.99 & 0.01 & 0.99 \\
Interactions of inputs, Firm-Year & 0.11 & 0.89 & 0.07 & 0.93 \\
\bottomrule
\multicolumn{5}{p{\dimexpr\linewidth-2\tabcolsep}}{{Note: All rows use the same Compustat sample ($N = 190{,}388$, 1980--2016), cleaned following the DLEU (2020) protocol with support restrictions from De Ridder, Grassi \& Morzenti (2024). Cost of Goods Sold (COGS) is the flexible input throughout; SG\&A is excluded to match the DLEU specification. We report the $R^2$ from regressions of form $\ln s^j_{it} = \alpha + \beta\ln\hat{\gamma}^j_{it}+\varepsilon_{it}$ in the Output Elasticity columns: the share of revenue-share variation linearly explained by the estimated output elasticity. The Markup columns report $1-R^2$: the residual share, attributed to markups. Cost-share rows proxy output elasticities with median industry-year COGS cost shares under constant RTS.}}
\end{tabularx}
\end{table}

Start with the Cobb-Douglas benchmark from \citet{deloecker2020rise}, which estimates sector-year output elasticities. Adding industry-year fixed effects to the regression sets the output elasticity's explanatory power to exactly zero by construction: a single elasticity applies to every firm in a given industry-year, so all within-industry variation in revenue shares is attributed to markups. More striking is how little changes in the full sample. Time-varying elasticities can absorb some revenue-share variation, but only about 4 percent of it; the markup residual explains the remaining 96 percent.

Cost-share proxies under constant RTS (Section~\ref{s:flexible_modeling}) tell a starker story. Because median industry-year cost shares assign one elasticity per industry-year cell, they have zero within-cell variation by construction. In the full sample, cost shares explain at most 1 percent of revenue-share variation; the markup residual absorbs the rest.

A natural response is to estimate elasticities more flexibly. Translog production functions allow estimated coefficients to interact with firm-specific inputs, producing output elasticities that vary across firms and time. The bottom panel of Table \ref{table:R2} reports results using the \citet{GrassiMorzentiRidder2022Hitchhiker} toolbox. Even here, the pattern persists. A translog without cross-input interactions explains about 1 percent of revenue-share variation with output elasticities, whether in the full sample or within industry-year. Adding cross-input interactions raises the elasticity share to 11 percent in the full sample and 7 percent within industry-year---the highest of any specification, but the markup residual still explains the vast majority. The additional flexibility helps at the margin, but the residual claimant---the markup---still dominates.

At face value, these translog results seem consistent with output elasticities simply not varying much. But there is a data problem: identifying translog terms requires enough similar firms to estimate how input ratios affect output. Compustat has few firms in narrow industries, and its inputs---Cost of Goods Sold and book capital---are coarse. Whether the low $R^2$ for elasticities reflects a real finding or a data limitation is hard to say with these data alone.

These results illustrate a broader point: the more that estimation restricts output elasticity variation, the more that revenue-share variation is mechanically attributed to markups. The R\textsuperscript{2} provides a helpful diagnostic but does not by itself reveal how much output elasticity variation is truly present in the data. Greater transparency on this front should be a priority for the literature, and the $R^2$ diagnostic offers a simple, low-cost way to achieve it.\footnote{Because $\ln s = \ln\gamma - \ln\mu$ is an identity, the full variance decomposition includes a covariance term. For the specifications in Table~\ref{table:R2}, this term is small enough that the simple regression $R^2$ closely tracks the variance shares. The advantage of the regression is that it requires no additional computation beyond what applied researchers already produce.}

\subsection{Stress-testing market power}
A high value avenue for research is to stress-test production-based markup estimates. Three strategies are especially promising: comparing production-based markups against demand-based markups common in industrial organization; using simulated data---where the truth is known---to evaluate estimator performance; and, exploiting natural experiments to test whether markup estimates detect real shifts in competitive conditions. 

First, the industrial organization literature has long estimated markups from the demand side. In principle, both demand and production approaches measure the same object---the ratio of price to marginal cost---but they do so through different data, assumptions, and methods. Comparing them in settings where both can be applied offers a valuable methodology and reality check. When both approaches agree, we gain confidence that the empirical signal is real. When they diverge, the comparison can pinpoint what each method is and is not capturing. Such cross-method comparisons remain surprisingly rare, and that is a missed opportunity.

The demand approach derives markups from a model of consumer behavior and firm pricing, typically applied to a narrowly defined industry---the classic example is \citet{berry1995automobile}'s study of the automobile market. Researchers specify a structural demand system, recover price elasticities from observed market shares and prices, and use a conduct model (such as Bertrand competition) to back out markups. The approach enables detailed welfare and counterfactual analysis but requires rich market-level data, functional-form assumptions about preferences and competition, and can be sensitive to instrument selection.

Do demand and production approaches to markups agree in practice? The evidence is mixed. \citet{grieco2023evolution} study the U.S. automobile industry from 1980 to 2018 using both methods. Their demand-based estimates show markups declining substantially over this period, while production-based estimates from \citet{deloecker2020rise} point in a different direction, both in levels and in trends. The authors trace much of the divergence to demand-side improvements in product quality and variety that production-based methods may not fully capture, a revealing finding about what each approach actually measures. A more encouraging comparison comes from \citet{deloecker-scott:2016}'s study of the U.S. beer industry, where both methods deliver broadly similar average markups, at least for certain years and specifications. A broader comparison comes from \citet{dopper2025rising} who estimate demand-based markups across over 100 consumer product categories. They find evidence of rising markups for 2006--2019 that are consistent with \citet{deloecker2020rise} in direction though not in magnitude.


Second, a complementary strategy is to use simulated data---where the true markups are known---to evaluate how well production-based estimators perform. \citet{GrassiMorzentiRidder2022Hitchhiker} conduct Monte Carlo simulations of this kind. Their results suggest that, in their specific model environment, using revenue rather than quantity data leads to biased markup estimates, though they argue that trends and dispersion are better recovered than levels. As we discussed in Section \ref{s:econometrics}, however, the one available empirical test---comparing revenue-based and quantity-based markups in French manufacturing data---yields a correlation of just 0.3, well below what the simulations would suggest. Whether simulation-based optimism about trends and dispersion survives contact with real data remains an important open question. Controlled simulation exercises nonetheless help map out the conditions under which production-based estimation can be trusted.

Third, natural experiments that generate plausibly independent shifts in competitive conditions can help test whether markup estimates detect the expected response. \citet{miller2017understanding} use the MillerCoors joint venture as such a test, showing that demand-based markup estimates correctly pick up the resulting price increases. \citet{carrillo2023misallocation} take a different approach: quasi-random variation in public procurement contracts for construction services in Ecuador identifies features of the marginal product distribution, allowing tests for misallocation and estimates of welfare losses. \citet{MajerovitzHughes_Experiments} go further, using a randomized experiment providing grants to Sri Lankan microenterprises to estimate the dispersion of marginal products---and hence the costs of misallocation---without imposing any production function assumptions. These settings not only help assess production-based methods but, when combined with structural estimation, can generate insights about firm behavior under imperfect competition that reduced-form approaches alone cannot deliver.

Stress testing should become a standard component of markup research, but this does not mean every paper must do everything. The production approach already demands substantial expertise across multiple fields, as this review has illustrated, and research benefits from specialization. Some researchers will push the frontier of measurement; others will focus on assessment through demand modeling, experimental design, or simulation. What we advocate is a collective effort: studies with rich data should compare multiple approaches; papers introducing new methods should demonstrate performance in simulations; empirical applications should seek quasi-experimental corroboration where available. Through this division of labor, the literature can build confidence in its measurements of market power and develop a clearer understanding of when and why different methods diverge.

\subsection{Aggregation: Macro implications of micro markups}\label{s:Aggregation}
Despite the conceptual and empirical challenges outlined earlier, firm-level markup estimates can meaningfully inform our understanding of productivity, welfare, and resilience to shocks. We highlight three research questions to guide the agenda for bridging micro measurement and macro modeling: What economic forces underlie measured markups? How much firm-level heterogeneity is needed to explain macroeconomic outcomes? And how do markups interact with the economy's network of input-output linkages?

First, knowing that markups exist is not enough; for policy analysis, we need to know why firms charge them. Markups can arise from very different sources: consumer demand elasticities, strategic pricing in oligopoly markets, the need to recover fixed costs, search frictions \citep{Menzio2025}, or increasing returns to scale. The production approach, by design, recovers equilibrium markups without taking a stand on what generates them. This makes it a powerful tool for documenting patterns in the data, but leaves economists without guidance on how to embed markups into structural models where the source of markup rents determines the welfare and growth implications.

Conflating different sources of markups can lead to misleading policy conclusions. Markups can represent pure economic profits that reduce efficiency and invite antitrust scrutiny. Or they can be a necessary byproduct of cost recovery: firms with high fixed costs or increasing returns must charge prices above marginal cost just to break even. If we treat all production-based markups as stemming from a single source---say, barriers to entry---when they partly reflect fixed costs or scale economies, our policy prescriptions can be misleading. Making progress requires treating production parameters themselves, not just markup estimates, as central objects of interest.\footnote{Indeed, in the 1990s production-approach literature, markups and returns to scale were treated as equally relevant. \citet{basu1997returns} emphasized returns to scale because of the focus of some models at the time, but their estimates mapped to markups via equation \eqref{e:markups_RTS}.}

Future research can help quantify the relative prevalence of ``good'' and ``bad'' markups from the long-standing macro-IO debates. Some markups reflect entry barriers or inefficient conduct that depresses output and growth \citep{bresnahan1989empirical, berry2019increasing, covarrubias2020good}. Others reflect Schumpeterian incentives: \citet{AghionBloomBlundell2005} find an inverted-U relationship between competition and innovation, with markups balancing positive incentive effects against negative efficiency costs (see \citealt{Gilbert2006} for a review). \citet{KletteKortum2004} and \citet{peters2020heterogeneous} model markups as the return to innovation. More recently, \citet{autor2020fall} find evidence consistent with superstar firms whose high markups reflect efficient scale and innovation. By contrast, \citet{Aghion2023falling_growth} argue that comparable markups can hinder growth when they stem from process inefficiencies and R\&D misallocation, that is, ``bad'' rents. Quantifying how prevalent each type is would help researchers understand how to rationalize production-based estimates in macroeconomic models.\footnote{\citet{Aghion2023falling_growth} and \citet{DeRidder2024} argue that good rents can morph into bad rents as technological winners erect barriers to entry, ultimately reducing innovation and growth.}

Second, how much heterogeneity matters for macro outcomes is a central open question. In other words: do we need the full distribution of firm-level markups to draw reliable macro conclusions, or does a representative firm get us close enough? On the household side, macroeconomists debate whether simplified two-agent New Keynesian (TANK) models capture enough heterogeneity, or whether richer heterogeneous-agent New Keynesian (HANK) models are required despite being harder to work with \citep{GaliLopezValles2007, KaplanMollViolante2018}. An analogous question arises for firms, and tractability pushes modelers toward parsimony; but ignoring important dimensions of heterogeneity risks distorting key counterfactuals.

Economic theory has not yet resolved how to aggregate firm-level markups into representative macro models. Standard business-cycle and growth models typically rely on a representative firm producing value added from capital and labor, without intermediate inputs; mapping firm-level gross-output markups into that framework is not straightforward. \citet{RotembergWoodford1995, RotembergWoodford1999} show that strong assumptions are needed to translate gross-output production with market power into value-added production. \citet{basu1997returns} argue that the right way to aggregate firm-level elasticities and markups depends on model specifics; for example, they provide a stylized example where the answer depends on whether marginal inputs are allocated proportionally to average inputs over the business cycle. 
No general answer yet exists for how much micro heterogeneity needs to be preserved when calibrating representative-firm models.\footnote{Growth accounting has long emphasized that microeconomic heterogeneity can explain macro aggregates. \citet{basu1997returns, basu2002aggregate} extend the accounting in \citet{JorgensonGollopFraumeni1987} to allow for heterogeneity in markups and returns to scale. The misallocation literature also emphasizes micro heterogeneity; \citet{baqaee2023micro} review the more recent literature, including \cite{baqaee2020productivity}.}

Ongoing empirical debates underscore why this heterogeneity question is so important. Revenue shares vary widely across firms but not randomly: Large, expanding firms tend to have low labor shares, with potentially important macro consequences \citep{autor2020fall, KehrigVincent2021}. \citet{VanReenen2018} documents increasing differences between firms in sales, productivity, and wages; when the measured aggregate markup increases, this increase is driven largely by reallocation toward larger, more productive firms. In rationalizing these patterns, workhorse models often feature firm heterogeneity in productivity \citep*{ericson1995markov, hopenhayn1992entry, melitz2003impact} and demand elasticities \citep*{AtkesonBurstein2008}, but typically assume common production technologies within industries. If empirical evidence of heterogeneous output elasticities proves robust, existing modeling frameworks will need to evolve. The choice of how to aggregate also matters: \citet{edmond2023how} show sales-weighted markups rise much more steeply than cost-weighted ones, with the gap reflecting allocative inefficiency from dispersion.\footnote{Markup heterogeneity need not imply misallocation. \citet{BornsteinPeter2025} show that under nonlinear pricing, markup heterogeneity across firms is an equilibrium feature that does not reflect inefficiency.}

Third, more research should help evaluate how the economy's network structure shapes the welfare consequences of markups. A long tradition of network analysis---from \citet{Leontief1941} to work reviewed by \citet{CarvalhoTahbazSalehi2019} and \citet{baqaee2023micro}---shows that intermediate input flows can amplify or attenuate distortions, so the location and propagation of markups through the network are as important as their average level.

Notably, upstream markups are likely more costly to the economy than downstream ones because they distort production directly rather than simply redistributing income. A markup on goods sold to final consumers may redistribute income from consumers to firms, with limited real effects, especially if labor is supplied inelastically. A markup on intermediate inputs sold to other producers acts more like a tax on production, distorting input choices and depressing aggregate productivity: it pushes the economy inside its production-possibilities frontier and violates the \citet{DiamondMirrlees1971} production-efficiency theorem. \citet{basu1995intermediate}, \citet{basu2002aggregate}, \citet{BigioLaO2020}, and \citet{baqaee2020productivity} formalize these effects.

Whether markups actually generate misallocation also depends critically on how firms contract with one another. When firms transact on spot markets, markups compound along supply chains through ``double marginalization,'' magnifying inefficiencies at each stage. But when upstream and downstream firms bargain directly---as in many long-term supplier relationships---markups need not generate misallocation even when they appear large. The macro consequences of markup heterogeneity therefore cannot be read off from markup estimates alone; they depend on network position and the institutional arrangements through which firms trade. Future research should clarify when markups primarily redistribute income, when they distort production, and how to incorporate these distinctions into tractable macro models.

\subsection{Takeaways for practice}
The preceding discussion points to three research priorities, each at a different stage of development.

First, researchers should report the $R^2$ decomposition of revenue-share variation into markup and output-elasticity components; if the markup explains nearly all the variation, that is a warning that the specification could be leaving too little room for technology. This diagnostic is available now and costs nothing beyond what applied researchers already compute.

Second, where the setting allows, production-based markups should be tested against outside evidence---demand-based estimates, simulations with known truth, or quasi-experimental shifts in competitive conditions. Such comparisons remain surprisingly rare, and expanding them is the most promising near-term path to putting the literature on firmer empirical footing.

Third, aggregation from firm to macro requires more careful attention to when markup heterogeneity reflects misallocation versus equilibrium dispersion, and to how network position shapes the welfare consequences of markups. This is the least resolved of the three and the most important for macroeconomists who want to use production-based estimates in general-equilibrium models.

Progress on all three fronts---transparency, testing, and aggregation---requires confronting the same fundamental challenge that runs through this entire review: production-based markups are residuals, and residuals are only as informative as the models, data, and econometrics that generate them.


\section{Conclusion}\label{s:conclusion}


The production approach to markups is both powerful and fragile. Its power comes from its minimal structure. With cost minimization and a flexible input, all we need is the input's share in revenue and its output elasticity to back out a markup. Its fragility comes from the same source. Because markups are residuals, whatever the model, the data, or the econometrics fail to capture gets loaded onto the markup. Researchers make a series of choices, each with its own risks.

For markup estimation, \cite{HallBPEA1986, Hall1988relation, HallR90} marked an important shift from ``craft to mass production'' \citep{Basu_Hall_remarks_2024}. Craft methods in the IO literature focus on careful studies of narrow industries with rich structural detail. Macroeconomists and trade economists need scale. Hall's production approach and the large literature applying his insights allow such scale.

Nevertheless, mass production in manufacturing requires much more than just a clear factory blueprint. For the factory to produce, it requires sourcing of reliable and high-quality inputs; it requires local institutional knowledge to acquire the needed permits and address planning commissions; and much besides. Similarly, applying the blueprint for production-based markups laid out in Section \ref{s:new_literature} is not just an off-the-shelf exercise. This review has highlighted many of the challenges that need to be acknowledged and, ideally, addressed.

Production-based markup estimates should not be taken at face value. The disagreements across studies are not minor: using the same data, different inputs imply opposite trends, different production functions yield markedly different levels, and different estimation methods produce divergent results. These disagreements reflect non-markup frictions, specification choices, data limitations, and econometric pitfalls---and they signal that the production approach's assumptions are difficult to satisfy in practice. More systematic attention to the checks developed in the preceding sections would sharpen the evidence on market power and reduce the risk that conclusions rest on artifacts of specification or measurement. To move forward, we need transparency, testing, robustness, and better data.

What should economists do with the evidence? We should acknowledge how much remains uncertain, while pressing ahead on both theory and measurement. On the theoretical side, there is no reason to wait. Markups shape resource allocation and welfare. They interact with misallocation, innovation, network structure, and cyclical dynamics. They also shape how shocks propagate across firms and industries. But we also need evidence on the structures that actually drive markups, in order to guide theorizing. The way forward is to use the approach honestly: report what the residual absorbs, test estimates against outside evidence, and treat specification choices as consequential rather than interchangeable. Measurement leaves us with residuals. The challenge is to combine institutional knowledge and economic theory to show what those residuals mean for growth, welfare, and policy.

\vspace{2em}
{\small
\setlength{\bibsep}{6pt plus 1pt minus 1pt}
\bibliography{bibliography}
}

\end{document}